\documentclass[pdflatex,sn-vancouver-num,referee]{sn-jnl}

\usepackage{graphicx}%
\usepackage{multirow}%
\usepackage{amsmath,amssymb,amsfonts}%
\usepackage{amsthm}%
\usepackage{mathrsfs}%
\usepackage[title]{appendix}%
\usepackage{xcolor}%
\usepackage{textcomp}%
\usepackage{manyfoot}%
\usepackage{makecell}
\usepackage{booktabs}%
\usepackage{algorithm}%
\usepackage{algorithmicx}%
\usepackage{algpseudocode}%
\usepackage{listings}%
\usepackage{natbib}
\usepackage{verbatim}
\usepackage{lineno}
\usepackage{hyperref}
\hypersetup{
citebordercolor=white,
linkbordercolor=white,
filebordercolor=white,
urlbordercolor=white,
linkcolor=black}
\newcommand{\Rem}{magnetic Reynolds number }
\newcommand{\Remp}{magnetic Reynolds number}
\newcommand{\rcr}{$\frac{r_c}{r}$ }
\newcommand{\Xw}{$X_w$ }
\newcommand{\Xs}{$X_{S,0}$ }
\newcommand{\rcrp}{$\frac{r_c}{r}$}
\newcommand{\Xwp}{$X_w$}
\newcommand{\Xsp}{$X_{S,0}$}
\newcommand{\foxp}{\textit{f}O$_2$}
\newcommand{\fox}{\textit{f}O$_2$ }
\newcommand{\mm}[1]{\mathrm{#1}} 

\begin{document}
\title{Dynamo generation reveals redox conditions during formation of differentiated planetesimals} 


\author*[1,2]{\fnm{Hannah R.} \sur{Sanderson}}\email{hannah.sanderson@visiting.ox.ac.uk}

\author[1]{\fnm{James F. J.} \sur{Bryson}}

\author[1]{\fnm{Claire I. O.} \sur{Nichols}}

\affil*[1]{\orgdiv{Department of Earth Sciences}, \orgname{University of Oxford}, \orgaddress{\street{South Parks Road}, \city{Oxford}, \postcode{OX1 3AN}, \country{UK}}}

\affil[2]{\orgdiv{Centre for Planetary Habitability (PHAB), Department of Geosciences}, \orgname{University of Oslo}, \orgaddress{\street{P.O. Box 1028 Blindern}, \city{Oslo}, \postcode{NO-0315}, \country{Norway}}}



\abstract{
In the early Solar System, an isotopic dichotomy existed between non-carbonaceous (NC) and carbonaceous (CC) planetesimals. Depending on the formation location of these planetesimals relative to condensation lines in the protoplanetary disk, NC and CC differentiated planetesimals could have had distinct redox states and water contents.  However, the extent of these differences and the resulting accretion environments of NC and CC planetesimals are debated. Here, we use thermal evolution and dynamo generation models to explore the effect of planetesimal core size, a proxy for redox state, and mantle water content on planetesimal dynamo generation. We find that combinations of core size and water content consistent with different formation scenarios produce planetesimals with stark contrasts in both magnetic field strength and duration. By comparing our models to existing paleomagnetic data for NC planetesimals, we suggest these bodies formed with a small amount of water-ice and degassed efficiently during differentiation. Future paleomagnetic measurements could determine whether CC planetesimals degassed as efficiently as NC planetesimals and the number of planetesimal formation regions in the NC reservoir. Overall, we demonstrate that meteorite paleomagnetism combined with dynamo generation models provides novel insight into the accretion environments of planetesimals and the evolution of their water contents.}




\maketitle

\addtocontents{toc}{\protect\setcounter{tocdepth}{0}} 
\section*{Introduction}\label{intro} 
A suite of measurements over the last $\sim$15 years have found that planetesimals formed in two isotopically distinct reservoirs: non-carbonaceous (NC) in the inner Solar System and carbonaceous (CC) in the outer Solar System \citep{warren_stable-isotopic_2011,bermingham_nc-cc_2020,kruijer_great_2020}. Samples from CC chondrites, which come from undifferentiated planetesimals in the CC reservoir, have higher matrix water contents and can be more oxidized than their NC counterparts \citep{alexander_quantitative_2019,alexander_quantitative_2019-1,amano_updating_2025}. Whether there were also differences in planetesimal redox state and water content between differentiated planetesimals in these reservoirs is an area of active debate \citep[e.g.,][]{newcombe_degassing_2023,grewal_accretion_2024,hellmann_hf-w_2024,spitzer_comparison_2025}. Differences in water content and redox state between these differentiated planetesimal populations would trace redox gradients in the disk and the locations of planetesimal formation \citep{grewal_accretion_2024}. Differentiated and undifferentiated planetesimals from both these reservoirs were the building blocks of the terrestrial planets, including the Earth \citep{dauphas_bayesian_2024,nimmo_mechanisms_2024}. Therefore, determining the formation environments of differentiated planetesimals in these reservoirs and their resulting compositions is key to advancing our understanding of protoplanetary disk evolution and the formation of the terrestrial planets. 

\begin{figure}
    \centering
    \includegraphics[width=1\linewidth]{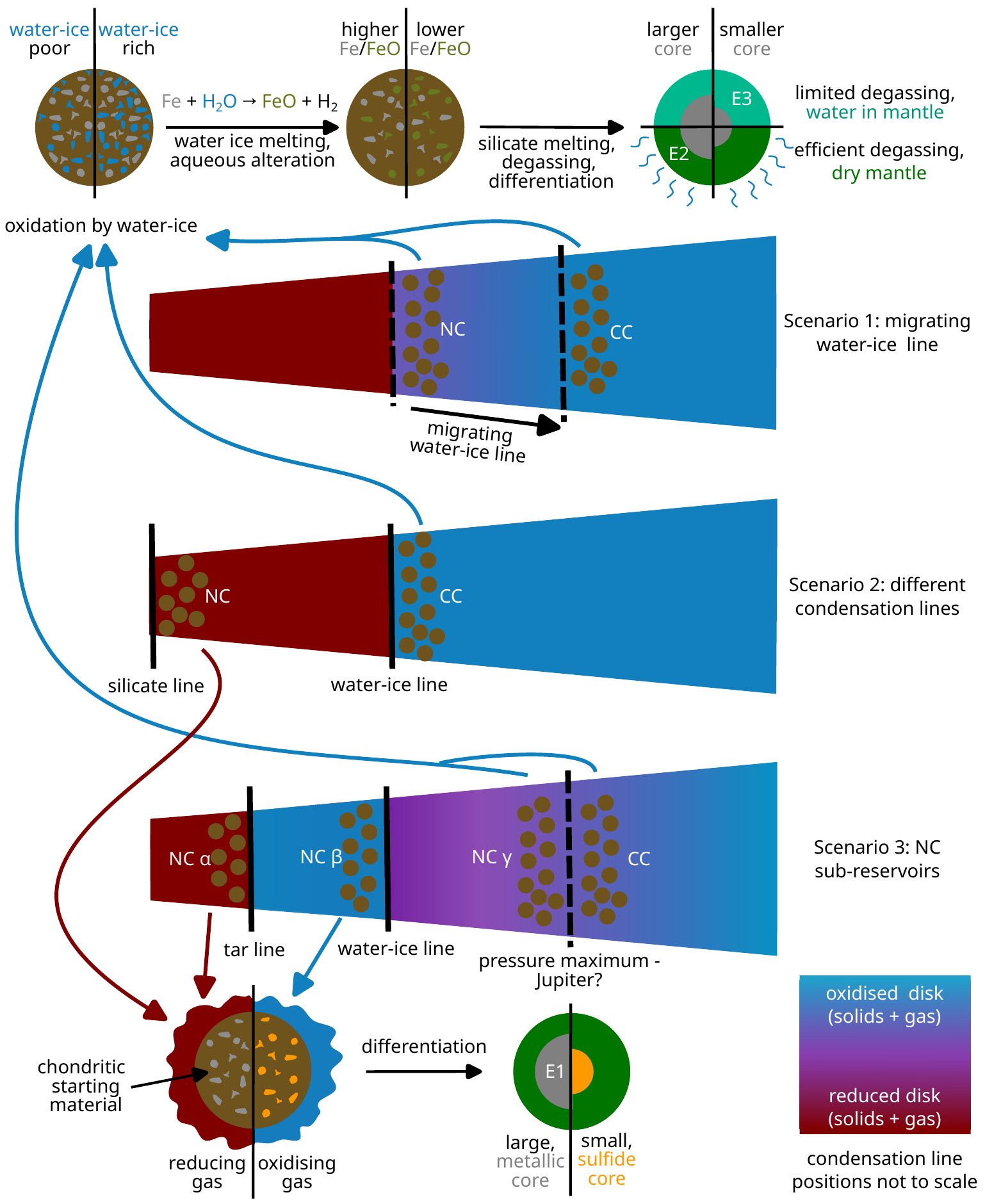}
    \caption{Illustration of possible disk structures, planetesimal accretion environments and differentiation pathways that result in different core sizes and mantle water contents in nominally anhydrous minerals. Scenario 1: migrating water-ice line \citep{drazkowska_planetesimal_2018,lichtenberg_bifurcation_2021}. Scenario 2: different condensation lines for NC and CC \citep{morbidelli_contemporary_2022,izidoro_planetesimal_2022}. Scenario 3: NC sub-reservoirs \citep{bermingham_nc-cc_2020,broadley_origin_2022,bryson_collective_2026}. E1, E2 and E3 denote the three planetesimal endmembers explored further in Figure \ref{fig:mag-main}. The colour of the disk indicates the relative degree of oxidation or reduction of planetesimals accreted in that region. The absolute position of the condensation lines is not to scale.  Outside the water-ice condensation line, water-ice is the dominant oxidising agent and the degree of oxidation depends on the amount of accreted water-ice. The degree of water retained in nominally anhydrous minerals in the mantle after differentiation depends on the efficiency of degassing Inside the water-ice condensation line, planetesimal oxidation is controlled by the absence of water ice (NC planetesimals in Scenario 2) and by the surrounding nebula gas. In the NC sub-reservoir hypothesis, the water vapour and solid organics between the tar and water-ice line produces an oxidising gas with a low C/O ratio, while interior to the tar line organic phases vaporise producing a reducing gas with high C/O ratio. In Scenario 1 the NC-CC isotopic dichotomy results from the migration of the water-ice line within the disk, while in Scenario 2 it results from planetesimal accretion at different condensation lines. In Scenario 3, an additional pressure maximum within the disk, possibly due to the formation of Jupiter, divides the reservoirs. }
    \label{fig:schematic}
\end{figure}

A planetesimal's water content and oxidation state is linked to where it formed in the disk relative to condensation lines. Planetesimals have been proposed to form adjacent to condensation lines because the accumulation of solids in these regions can sufficiently increase density to trigger the streaming instability which begins planetesimal formation \citep{lichtenberg_bifurcation_2021,izidoro_planetesimal_2022}. Condensation lines relevant for planetesimal formation have been proposed to be the silicate sublimation line ($\sim1400$\,K), the tar line ($\sim400$\,K), and the water-ice line ($\sim$170\,K), outside which silicates, organics, and water, respectively, are solids \citep{izidoro_planetesimal_2022,lodders_jupiter_2004,nakano_precometary_2020,bermingham_nc-cc_2020}. For planetesimals outside the water-ice line, accreted water-ice was the dominant oxidising agent \citep{lauretta_action_2006,grossman_formation_2012,sutton_bulk_2017}. As a planetesimal heated up due to decay of radioactive $\rm^{26}Al$, water-ice melted and reacted with iron to form iron oxide (Figure \ref{fig:schematic}). Inside the water-ice line, planetesimals did not contain significant water-ice and the gas in the protoplanetary disk was the main oxidising/reducing agent. Between the water-ice line and the tar line, the presence of water in the gas phase lowered the C/O ratio relative to solar abundances producing an oxidising gas \citep{righter_redox_2016,bermingham_nc-cc_2020,bryson_collective_2026}. Inside the tar line, the gas had a solar C/O ratio and was reducing \citep{grossman_formation_2012,righter_redox_2016}.

Three different scenarios for planetesimal formation at condensation lines  have been proposed, which have different consequences for planetesimal oxidation state and initial water content (Figure \ref{fig:schematic}). In the `migrating water-ice line' scenario (Scenario 1, Figure \ref{fig:schematic}) \citep{drazkowska_planetesimal_2018,lichtenberg_bifurcation_2021}, NC and CC planetesimals formed sequentially at the water-ice line as the protoplanetary disk evolved and this condensation line migrated. Since both NC and CC planetesimals would have formed at the water-ice line, NC and CC planetesimals would have had similar oxidation states. In the `different condensation line' scenario (Scenario 2, Figure \ref{fig:schematic}) , NC and CC planetesimals formed simultaneously at the silicate condensation line and the water-ice line, respectively \citep{izidoro_planetesimal_2022,morbidelli_contemporary_2022}. Here, NC planetesimals would have been free from water-ice and more reduced than their CC water-ice-bearing counterparts. Finally, in the `NC sub-reservoirs' scenario (Scenario 3, Figure \ref{fig:schematic}), some NC and all CC planetesimals formed simultaneously outside the water-ice line but were separated by a pressure maximum, possibly caused by Jupiter \citep{bermingham_nc-cc_2020,bryson_collective_2026}. These NC planetesimals would have accreted water-ice and had a similar oxidation state to CC planetesimals. At the same time, some NC planetesimals also would have formed between the tar line and water-ice line, and inside the tar-line. This would have resulted in two further populations of NC planetesimals: one more reduced than CC planetesimals and one more oxidised than CC planetesimals \citep{bermingham_nc-cc_2020,bryson_collective_2026}. 

\begin{figure}
    \centering
    \includegraphics[width=1\linewidth]{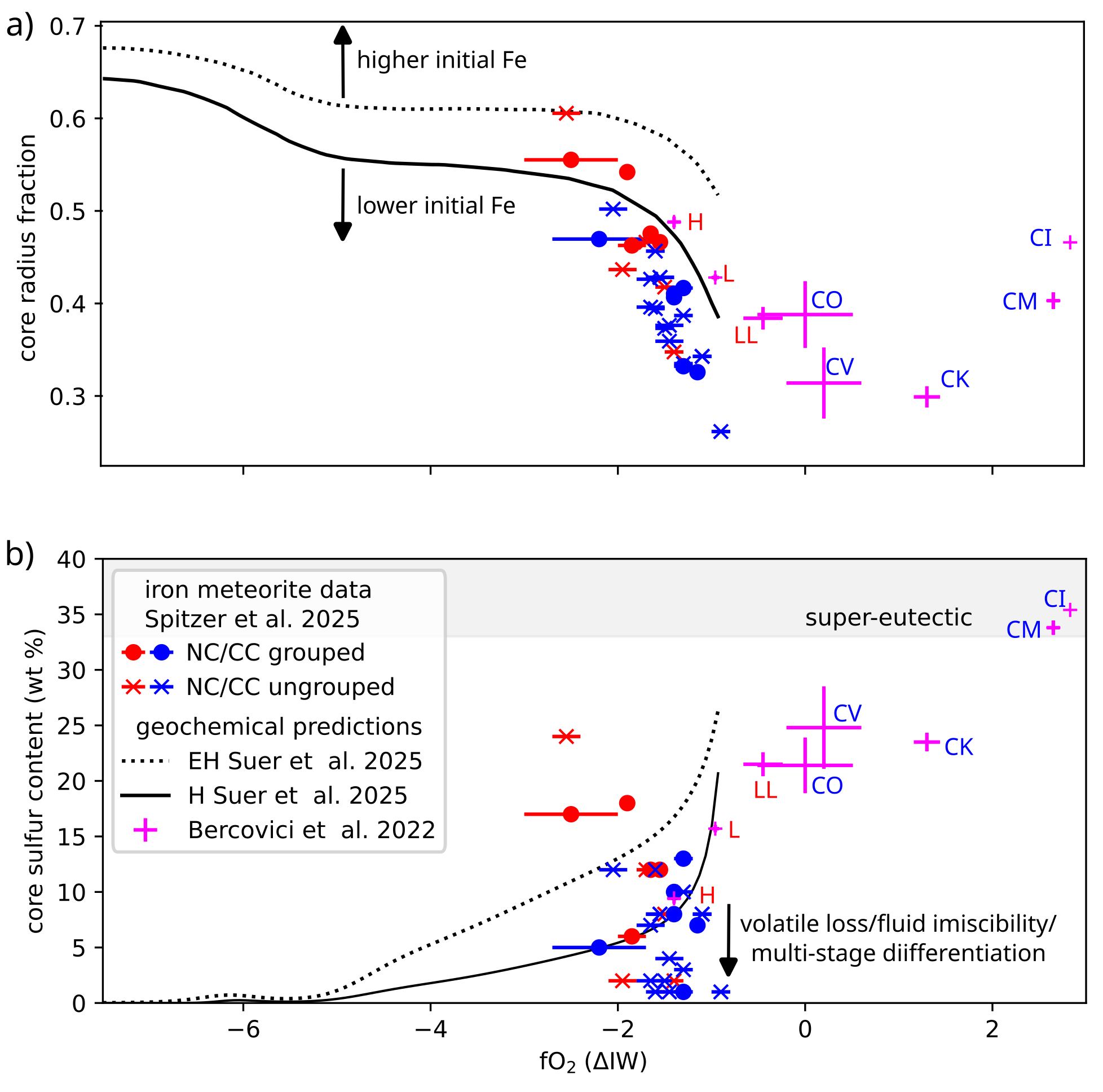}
    \caption{a) Core radius fraction and b) core sulfur content as a function of redox state quantified by oxygen fugacity, \foxp, relative to the iron-w{\"u}stite (IW) buffer, based on data from iron meteorites \citep{spitzer_comparison_2025} and predicted by geochemical models for chondritic starting compositions \citep{bercovici_effects_2022,suer_formation_2025}. The colour of the labels for the pink crosses \citep{bercovici_effects_2022} indicates whether the chondrite is NC (red) or CC (blue). For the iron meteorites, the average of the oxygen fugacities calculated from Fe/Co and Fe/Ni were plotted and the total width of the error bars is equal to the difference between these two fugacities. There are no vertical errorbars because \citet{spitzer_comparison_2025} did not provide errors on core mass fraction or core sulfur content from which these radii were calculated. For more detail on the calculation of core radius fractions from the iron meteorite data see Section \ref{supp-rcr}. Both oxygen fugacity and initial composition affect core size \citep{bercovici_effects_2022,suer_formation_2025}, which may contribute to the scatter along the trend in a). In b), the iron meteorite data for sulfur content as a function of fugacity does not match the trends predicted by geochemical models probably due to some combination of volatile loss during differentiation, \citep{hirschmann_early_2021}, fluid immiscibilty \citep{bercovici_effects_2022,bromiley_effects_2026}, and multi-stage differentiation \citep{grewal_protracted_2025}}
    \label{fig:fo2-rcr}
\end{figure}

Two ways these hypotheses for planetesimal formation have previously been tested are by determining NC and CC planetesimal accretion times \citep{spitzer_nucleosynthetic_2021,hellmann_hf-w_2024,bryson_collective_2026} and the redox states of magmatic iron meteorites, which sample the cores of differentiated planetesimals \citep{hilton_chemical_2022,grewal_accretion_2024,spitzer_comparison_2025}. However, these measurements do not support just one formation hypothesis. Similar NC and CC planetesimal accretion times supports either the `different condensation lines' or `NC sub-reservoirs' hypotheses \citep{spitzer_nucleosynthetic_2021,bryson_collective_2026}, while the overlapping redox states of NC and CC magmatic iron meteorites (Figure \ref{fig:fo2-rcr}) leans towards the `migrating water-ice line' or `NC sub-reservoirs' hypothesis. Additionally, these measurements are subject to several sources of uncertainty, such as  assumptions in fractional crystallisation models \citep{grewal_accretion_2024,spitzer_comparison_2025}; and initial elemental abundance ratios used in radioisotope dating \citep{hellmann_hf-w_2024} (for further details see the Supplementary Materials). 

Formation environments have also been tested using achondrite water contents \citep[e.g.,][]{newcombe_degassing_2023,harries_upper_2023,peterson_h_2023,stephant_hydrogen_2023,rider-stokes_evidence_2024}, but this is complicated by water loss during planetesimal heating and differentiation. During differentiation, most water is degassed due to the dehydration of hydrous minerals at high temperature \citep[$\sim1200$\,K;][]{lichtenberg_bifurcation_2021}, but a small amount of water may be retained in nominally anhydrous minerals (NAMs) \citep{fu_silicate_2017}. We consider two possibilities for the relationship between differentiated planetesimal water contents and their formation environment. The first possibility is that some water retention is possible, so a differentiated planetesimal's water-content in NAMs is related to the water-content of the undifferentiated (chondritic) precursor \citep[e.g.,][]{rider-stokes_evidence_2024}. In this case, an achondrite with negligible water originated from a planetesimal that accreted in water-ice free conditions. The second possibility is that planetesimals degas very efficiently, so differentiated planetesimal water contents are independent of their chondritic precursor's composition \citep[e.g.,][]{newcombe_degassing_2023,peterson_h-poor_2024,peterson_reconstruction_2025}. In this case, an achondrite with negligible water could have originated from a planetesimal that accreted either in water-ice free conditions or with water-ice and then degassed efficiently (Figure \ref{fig:schematic}). These two possibilities can be tested by comparing water in NAMs in CC achondrites, i.e., bodies that definitely formed with water-ice, with water in NAMs in NC achondrites, i.e., bodies that may have formed with water-ice. However, these measurements are challenging and studies report a wide range of parent body water contents \citep{newcombe_degassing_2023,peterson_reconstruction_2025}, some of which are contradictory within a given meteorite group  \citep[e.g., angrites;][]{sarafian_early_2017,deligny_origin_2021,rider-stokes_evidence_2024} (for more details see the Supplementary Materials). Altogether, based on existing approaches, the formation environments of planetesimals and the extent to which they can retain water during differentiation is unclear. 

In this work, we demonstrate a new approach to test differences in oxidation state and water content between planetesimal reservoirs by using differentiated planetesimals' magnetic and thermal histories. These histories can be recovered from paleomagnetic remanences in meteorites \citep[for overviews see][]{weiss_paleomagnetic_2010,dodds_thermal_2021,sanderson_early_2024} and thermochronometers \citep[e.g.,][]{neumann_modeling_2018,bryson_paleomagnetic_2019}. We use a refined thermal evolution and dynamo generation model, adapted from \citet{sanderson_unlocking_2025}, to explore the effects of mantle water content and oxidation state on thermal evolution and dynamo generation (see \nameref{met}). We incorporate increases in mantle water content by lowering the mantle viscosity and the mantle solidus. We implement changes in oxidation state, which can be quantified by oxygen fugacity, \foxp, through changes in core radius fraction. For a given composition, the core radius within a more oxidised planetesimal (up to the iron-w{\"{u}}stite buffer, \fox=IW) will be a smaller fraction of the total planetesimal radius because less Fe is available to contribute to core formation (Figure \ref{fig:fo2-rcr}a). We do not include changes in core sulfur content with oxidation state because there is no clear trend between oxygen fugacity and calculated core sulfur content in the iron meteorite data (Figure \ref{fig:fo2-rcr}b) and core sulfur contents can be altered from expected abundances during and after differentiation \citep{bercovici_effects_2022,hirschmann_early_2021,bromiley_effects_2026} (see Supplementary Materials). 

We compare predicted thermal and magnetic histories for three planetesimal endmember compositions that could result from a range of accretion environments and water-loss histories (E1---E3, Figure \ref{fig:schematic}). Endmember 1 accreted without water and was reduced by nebular gas; Endmember 2 accreted with a small amount of water-ice but lost all its water during differentiation; and Endmember 3 accreted with more water ice than Endmember 2 and retained a small amount of water in NAMs after differentiation. All three endmembers could represent NC planetesimals, depending on the planetesimal formation model, while Endmembers 2 and 3 could represent CC planetesimals, since they formed beyond the water-ice line. We use these endmembers to show that meteorite paleomagnetism can shed light on the redox states and water contents of differentiated planetesimals, while thermochronometric data cannot. This enables us to recover the accretion environments of some NC planetesimals and distinguish between the proposed scenarios for their formation.

\section*{Results}\label{res} 
\begin{figure}
    \centering
    \includegraphics[width=1\linewidth]{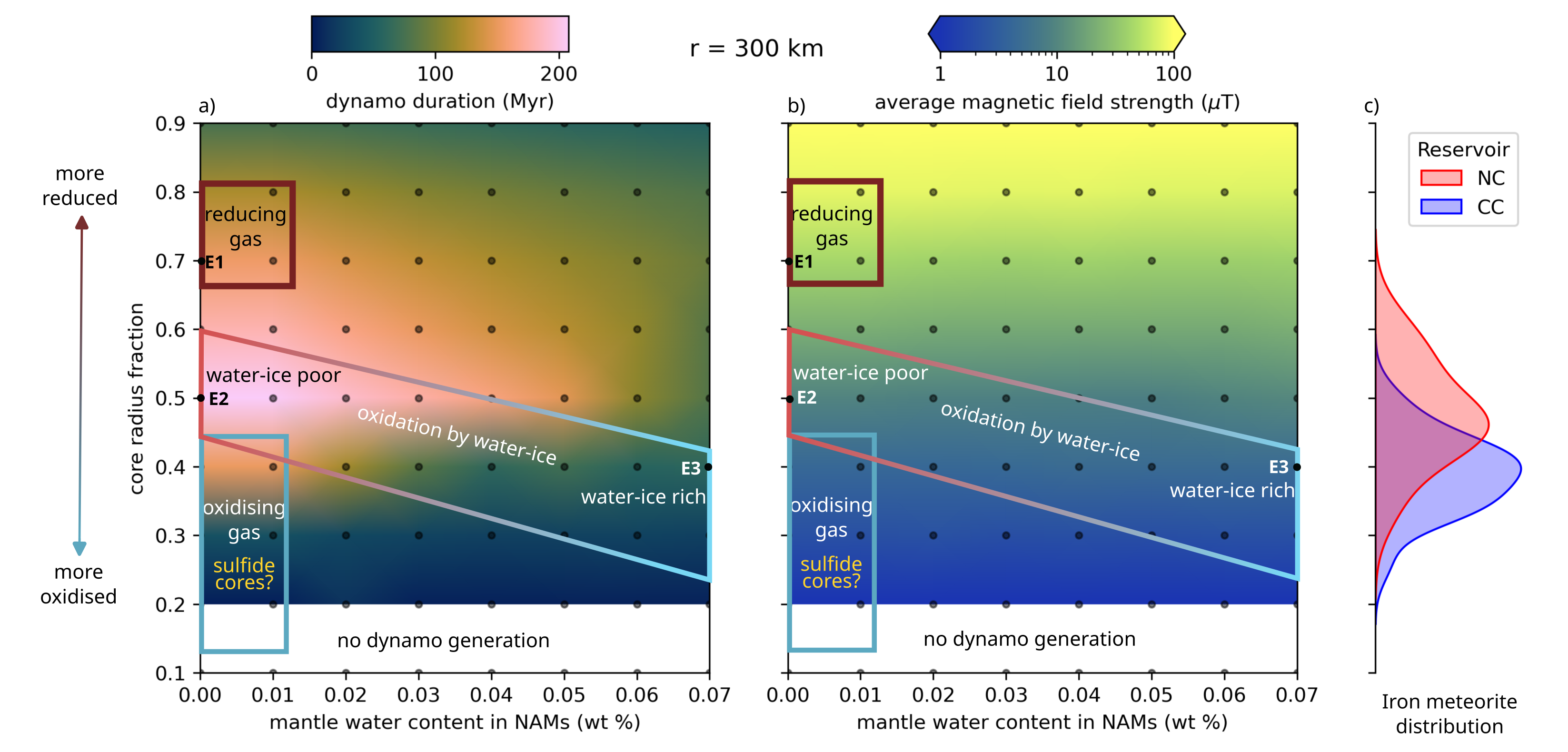}
    \caption{a) Total dynamo duration and b) average magnetic field strength as a function of mantle water content in nominally anhydrous minerals (NAMs) and core radius fraction for a 300\,km radius planetesimal. The white regions indicate parameter combinations that did not result in a dynamo. 0.0126\,wt\% water is the threshold above which water significantly changes the rheology and solidus (Section \ref{supp-xw}), so values below this are equivalent to 0\,wt\% water. For planetesimals with multiple epochs of dynamo generation, the duration of all epochs were summed to calculate the total dynamo duration. Boxes indicate possible regions of parameter space occupied by planetesimals from which formed in different regions in the disk and were exposed to different oxidising agents. Further justification of boundary positions is given in Section \ref{supp-redox}. Endmembers 1--3 explored in Figure \ref{fig:mag-main} are denoted by E1--E3. The possibility of sulfide cores is discussed in the Supplementary Materials. c) The distribution of core radius fractions within NC and CC iron meteorites based on data from \citet{spitzer_comparison_2025}. The area under each distribution is normalised to one. The core radius fractions of E2 and E3 are based on the mean core radius fractions of the NC and CC iron meteorites, respectively. The same figure for 100\,km and 500\,km radius planetesimals is shown in \ref{fig:regime100} and \ref{fig:regime500}, respectively.}
    \label{fig:regime300}
\end{figure}

We explore the combined effects of planetesimal core size and mantle water content (\ref{tab:params}) on dynamo duration and surface magnetic field strength (Figure \ref{fig:regime300}). To generate a magnetic field, a planetesimal must have a molten, metallic core in which sufficiently vigorous flow is being driven. This flow can be driven by thermal convection arising from superadiabatic core cooling and/or compositional convection arising from core solidification. 

Both core size and mantle water content affect dynamo duration (Figure \ref{fig:regime300}a). Dynamo duration initially increases with increasing core radius fraction due to the increase in core convective lengthscale, which decreases the rate of core cooling/solidification needed for dynamo generation \citep[see \nameref{met} and][]{sanderson_unlocking_2025}. Since absolute core size controls dynamo generation, the lower limit on core radius fraction is higher for smaller planetesimals (\ref{fig:regime100} and \ref{fig:regime500}). For large core radius fractions, dynamo duration becomes limited by rapid mantle and core cooling, which brings forward the time of complete core solidification. At intermediate core radius fractions, mantle water content becomes the limiting factor in dynamo duration. Addition of water lowers the mantle solidus \citep{katz_new_2003} and planetesimal peak temperature, which decreases the degree of cooling required before the onset of core solidification (\ref{fig:solidus-eta}). Addition of water also lowers the mantle viscosity \citep{hirth_water_1996,keller_volatiles_2017}, which increases the mantle, and hence core, cooling rate. Together the lower planetesimal peak temperature and lower mantle viscosity bring forward the time when the core reaches the eutectic composition and dynamo generation ends (see \nameref{met}). In contrast, we show that surface magnetic field strength only depends on core size (Figure \ref{fig:regime300}b). Larger cores produce stronger surface fields because magnetic field strength follows an inverse cube law so decreases rapidly with increasing distance from the CMB. 

\begin{figure}
    \centering
    \includegraphics[width=1\linewidth]{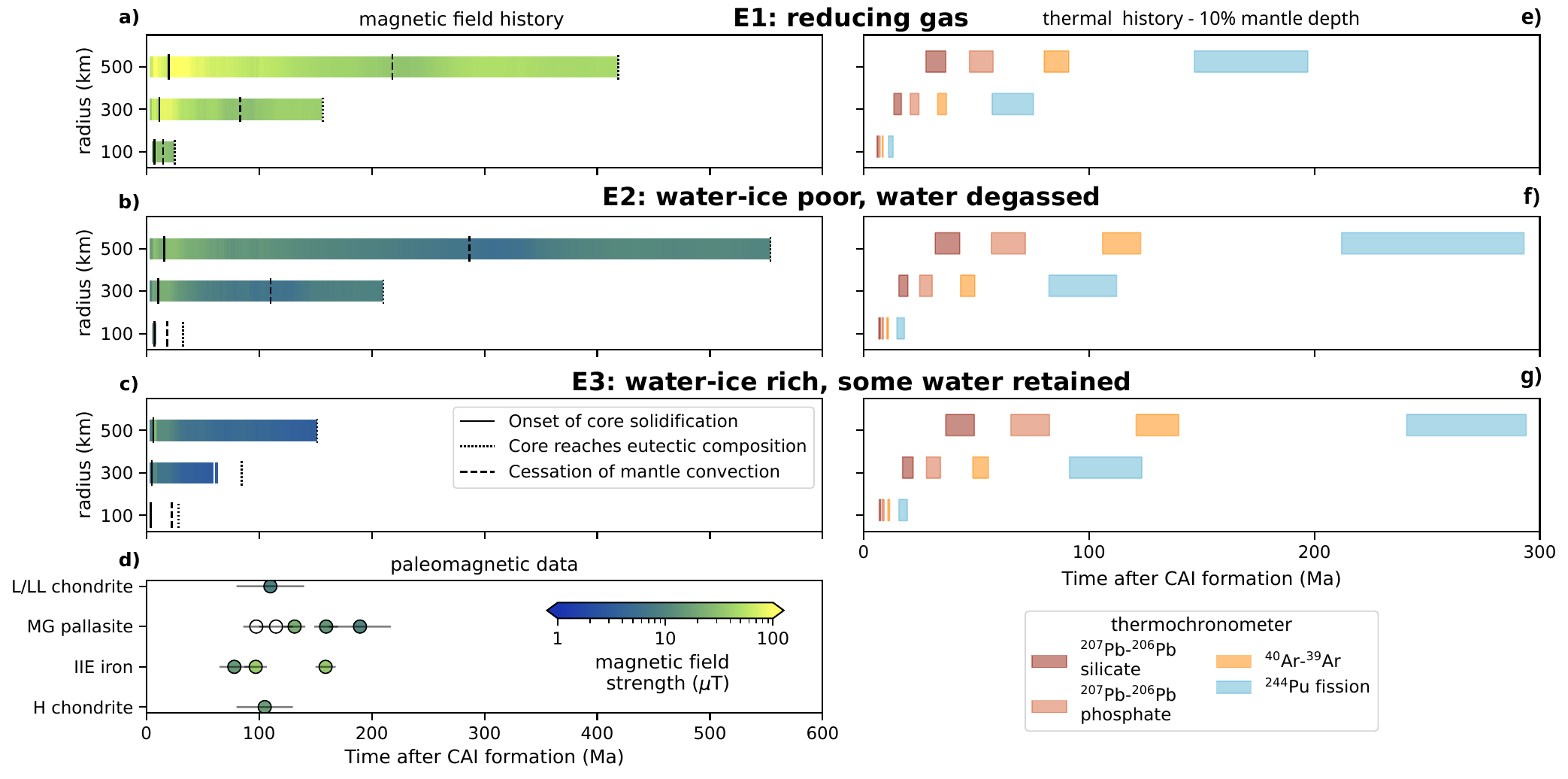}
    \caption{Dynamo strength and duration (a, b, c) and closure time for thermochronometers (e, f, g) for three planetesimal endmembers from Figure \ref{fig:regime300}. Endmember 1 (a, e): dry planetesimals reduced by nebula gas ($\frac{r_c}{r}=0.7,\: X_w=$0\,wt\%). Endmember 2 (b, f): planetesimals that accreted a small fraction of water-ice that degassed all their water during differentiation ($\frac{r_c}{r}=0.5,\: X_w=$0\,wt\%). Endmember 3 (c, g): planetesimals that accreted a lot of water-ice and retained some water in the mantle after differentiation ($\frac{r_c}{r}=0.4,\: X_w=$0.07\,wt\%). In the left column, filled bars indicate periods when a dynamo is active and the colour of the bar indicates dipole magnetic field strength at the surface. The black, vertical lines indicate the onset of core solidification (solid), when the core reaches the eutectic composition (dotted), and the cessation of mantle convection (dashed). d) Meteorite paleomagnetic data for meteorite groups with non-zero paleointensity $>50$\,Ma after CAI formation. All meteorites are from the NC reservoir. Circles are colour coded by the measured mean paleointensity on the same colour scale as the upper panels (Table \ref{tab:paleoint}). White circles indicate null paleointensities. All data comes from the following studies \citet{shah_long-lived_2017,maurel_meteorite_2020,maurel_long-lived_2021,tarduno_evidence_2012,bryson_long-lived_2015,nichols_pallasite_2016,nichols_time-resolved_2021,bryson_constraints_2019,wang_lifetime_2017}. In the right panel, the width of the box indicates the possible closure time recorded by thermochronometers that have previously been used to trace planetesimal thermal histories \citep{bryson_paleomagnetic_2019}. The width of the box corresponds to the range of closure times from the range in closure temperatures for a given isotope system. The ranges are: $^{207}$Pb-$^{206}$Pb in silicate 950--1150\,K; $^{207}$Pb-$^{206}$Pb in phosphate 700--800\,K; $^{40}$Ar-$^{39}$Ar 530--570\,K; $^{244}$Pu fission 365--415\,K.}
    \label{fig:mag-main}
\end{figure}

Across this parameter space of core radius fraction and mantle water content, both total dynamo duration and time-averaged surface magnetic field strength (excluding gaps in dynamo generation) vary by two orders of magnitude. This demonstrates that planetesimals that formed under different protoplanetary disk conditions and/or with different degassing histories (boxes on Figure \ref{fig:regime300}) will have very different magnetic histories; readily distinguishable using meteorite paleomagnetism. In Figure \ref{fig:mag-main}, we present thermal and magnetic field histories for 100--500\,km radius planetesimals for three endmembers with distinct oxidation states and water contents (E1--E3 in Figure \ref{fig:schematic} and \ref{fig:regime300}) and compare them to time-resolved paleomagnetic records. This allows us to account for the effects of planetesimal size on both the dynamo duration and magnetic field strength \citep[\ref{fig:regime100}, \ref{fig:regime500}, and][]{sanderson_early_2024}. 

We show that the planetesimal endmembers have starkly contrasting magnetic histories but similar thermal histories (Figure \ref{fig:mag-main}). Planetesimals reduced by nebula gas (E1) have magnetic field strengths that are one to two orders of magnitude stronger than planetesimals oxidised by water ice (E2 and E3), irrespective of planetesimal size, due to their larger fractional core radii. Planetesimals slightly oxidised by water-ice that degassed their water during differentiation (E2) generate dynamos for the longest time due to their dry, viscous, reasonably-thick mantles, which cool slowly. Differences in duration due to core radius fraction and water content dominate over differences due to planetesimal radius. For example, 500\,km radius planetesimals that were water-ice rich and did not degas efficiently (E3) generate dynamos for a shorter time than 300\,km radius planetesimals for E1 and E2. 

In contrast, we show that thermochronometers cannot distinguish between the same size planetesimal across different planetesimal endmembers. This is because the temperature ranges recorded by thermochronometers (350--1200\,K) are only reached in the conductive portion of the planetesimal, which starts at the planetesimal surface and extends to greater depths as the mantle cools. \citep[Figure \ref{fig:temp-profile};][]{henke_thermal_2012,bryson_paleomagnetic_2019}. The timescale for conductive cooling is proportional to the square of the length over which heat is conducted. Therefore, mantle thickness, which depends on planetesimal size, controls the cooling timescale and the closure time of thermochronometers. 

\section*{Discussion}\label{dis}
We calculate that planetesimal endmembers, corresponding to different accretion environments within the protoplanetary disk, have distinct magnetic histories. In contrast, differences in their thermal histories, as recorded by thermochronometers, are degenerate with differences in parent body size. Therefore, meteorite paleomagnetism is uniquely placed to reveal the redox conditions during formation and final water content of differentiated planetesimals. The consistent difference in magnetic field strength between the endmembers occurs $>$50\,Ma after CAI formation, so paleomagnetic data from after this time is the best proxy for the water content and oxidation state of planetesimals. Only paleomagnetic data from NC meteorites is currently available in this time range \citep[Figure \ref{fig:mag-main}d and][]{sanderson_early_2024}. The NC paleointensities are closest to those produced by the water-ice poor, efficiently degassed endmember (E2) except for the two youngest acquired remanences in the Main Group pallasites which are weaker and closer to the water-ice rich, water-bearing mantle endmember (E3). However, these values have very large uncertainties which span the paleointensity range of all three endmembers due to the unknown direction of the paleofield \citep{nichols_time-resolved_2021}. The extended duration of the fields within the paleomagnetic record can be produced by 300--500\,km radius reduced gas (E1) and water-ice poor, efficiently degassed (E2) planetesimals. Altogether, these intermediate strength and long-lived fields suggest that NC differentiated planetesimals were most similar to the water-ice poor endmember and formed either at a migrating water-ice condensation line (Scenario 1, Figure \ref{fig:schematic}) or in the outermost NC sub-reservoir (Scenario 3, Figure \ref{fig:schematic}). 

The match between the paleomagnetic data and Endmember 2, which has negligible mantle water content, supports previous evidence that planetesimals degas very efficiently during differentiation \citep{harries_upper_2023,newcombe_degassing_2023,grewal_implications_2025,peterson_reconstruction_2025}. However, current paleomagnetic data is only for NC achondrites and CC differentiated planetesimals may have degassed less efficiently. CC differentiated planetesimals may have retained more water in NAMs after differentiation compared to some NC differentiated planetesimals, because CC differentiated planetesimals may have accreted with a higher initial amount of water-ice. Two observations suggest this higher initial water-ice content in CC differentiated planetesimals compared to their NC counterparts. Firstly, the average core size of CC iron meteorite parent bodies is slightly smaller than NC iron meteorite parent bodies (Figure \ref{fig:regime300}). Secondly, CC chondrites have higher fractions of matrix, the water-ice bearing phase, than NC chondrites \citep{alexander_quantitative_2019,alexander_quantitative_2019-1,bryson_collective_2026}. Paleomagnetic measurements of CC achondrites could test whether a planetesimal's initial water content affected the extent to which it retained water in NAMs post-differentiation. If CC achondrites have paleomagnetic histories similar to Endmember 2, this would further support the efficient degassing hypothesis and suggest that the water content of planetesimal mantles is independent of the initial amount of water accreted. Conversely, if the paleomagnetic histories of CC achondrites are similar to Endmember 3, this would suggest that mantle water contents are affected by pre-differentiation water contents and some variation in water content between achondrites could be primordial. 

One possible target for paleomagnetic measurements are ungrouped rocky achondrites. However, those that have been dated cooled too early \citep[all formed $\leq$10\,Ma after CAI and cooled rapidly to below their closure temperature;][]{bouvier_new_2011,amelin_u-pb_2019,sanborn_carbonaceous_2019,hyde_detailed_2022,rider-stokes_rapid_2025} to record a remanence when the endmembers exhibit the starkest contrast. Other more promising targets are the Eagle Station pallasites and ungrouped stony-irons Bocaiuva, Mbosi, NWA 176, and Tucson \citep{malvin_bocaiuva--silicate-inclusion_1985,liu_northwest_2001,ruzicka_silicate-bearing_2014,spitzer_comparison_2025}. The mixture of metal and silicate in these meteorites suggest these meteorites originated from the mid-mantle within their parent bodies \citep{ruzicka_silicate-bearing_2014}. Therefore, they are more likely to have acquired a paleomagnetic remanence $>50$\,Ma after CAI formation. Additionally, these samples are analogous to two of the NC meteorite groups that have later remanences: the silicate-bearing IIE irons and the Main Group pallasites \citep{tarduno_evidence_2012,bryson_long-lived_2015,nichols_pallasite_2016,maurel_meteorite_2020,maurel_long-lived_2021,nichols_time-resolved_2021}. Therefore, they could provide a direct point of comparison between similar NC and CC bodies.

While existing NC paleomagnetic data supports planetesimal formation via either Scenario 1 or Scenario 3, discriminating between these scenarios requires additional observations. Paleomagnetic measurements on achondrites from the two potential NC sub-reservoirs inside the water condensation line (Scenario 3) would enable us to determine whether they formed in different reservoirs to NC achondrites with existing paleomagnetic measurements. Aubrites formed in a very reduced environment \citep{keil_enstatite_2010} and their parent body is thought to be large and differentiated \citep{cartier_large_2022,steenstra_geochemical_2020}. Therefore, they could record paleomagnetic histories similar to Endmember 1, which would suggest some planetesimals formed in a reduced NC sub-reservoir inside the tar line. Brachinites are thought to have formed in oxidising conditions \citep{righter_redox_2016,crossley_parent_2023}, so could sample the region between the tar and water-ice lines. Since they are primitive achondrites (i.e., they experienced a low degree of partial melting) they must have originated from the outermost layers of a partially differentiated body if they are found to record a magnetic field. Another option is angrites, with a recently dated angrite \citep{rider-stokes_impact_2024} being a promising candidate for paleomagnetic measurements because it formed in the time period when our planetesimal endmembers produce distinct magnetic histories. There is evidence for angrite formation in a very oxidizing environment, i.e., between the water-ice and tar lines, \citep{righter_redox_2016} and in a more reducing environment, i.e., just outside the water-ice line  \citep{steenstra_effect_2017,wang_lifetime_2017,kleine_chronology_2012,bell_appraising_2023}. Comparing angrite paleomagnetic data to Endmember 2 and future predictions of dynamo generation in sulfide cores could help determine the oxidation state of the angrite parent body, and hence the disk conditions in which it formed.

In summary, planetesimal dynamo generation is strongly controlled by mantle water content and core radius fraction (i.e., redox state). Planetesimals that formed under different redox conditions and retained different amounts of water after differentiation have stark contrasts in magnetic field strength and dynamo duration. Meanwhile, thermochronometers are minimally affected by planetesimal redox state and mantle water content and are instead controlled by planetesimal size. As a result, meteorite paleomagnetism combined with dynamo generation modelling provides a new, unique approach to assess redox conditions during planetesimal formation and hence test planetesimal formation hypotheses. Additionally, this method can constrain the amount of water retained by planetesimals during differentiation. Existing paleomagnetic measurements suggest that some NC planetesimals accreted in an environment with water ice and degassed efficiently during differentiation. Future measurements could reveal the number of formation environments in the NC reservoir and the extent of degassing of CC planetesimals. Altogether, this new approach promises to advance our understanding of the formation and evolution of planetary building blocks.

\clearpage
\section*{Methods}\label{met} 
\subsection*{Model summary}
We used a modified version of the planetesimal thermal evolution and dynamo generation model presented by \citet{sanderson_unlocking_2025} to investigate possible differences in the thermal and dynamo histories between NC and CC differentiated planetesimals. Our model is spherically symmetric and 1D. It tracks the thermal evolution of both the core and the mantle and determines the strength and duration of dynamo generation using scaling laws. Our model includes convection and conduction in both the core and the mantle. Mantle convection is assumed to be in the stagnant lid regime with stagnant, conductive boundary layers at the surface and CMB that thicken as the mantle cools and mantle viscosity increases. Mantle convection ceases when the combined thickness of these boundary layers is equal to the mantle thickness. 

In planetesimals, a dynamo can be generated by thermal convection, compositional convection or a combination of the two. Thermal convection occurs when the CMB heat flux is superadiabatic and compositional convection occurs when the core is solidifying. Our model assumes that the core is a sub-eutectic Fe-FeS alloy (see \nameref{met-xs}), undergoes perfect fractional crystallisation, and solidifies inwards due to the low CMB pressures in planetesimals \citep[for discussions of the core solidification mechanism see][]{sanderson_unlocking_2025}. In this scenario, core solidification drives convection due to the density contrast between solidified pure iron and sulfur enriched liquid, assuming complete partitioning of S into the liquid phase. Once the liquid part of the core reaches the eutectic composition (33\,wt\% S), compositional dynamo generation ceases because there is no longer a significant density contrast between the solid and liquid phases, which subsequently form with the same composition. 

Our model assesses whether dynamo generation is possible using the non-dimensional \Remp. When the \Rem is supercritical \citep[$>10$]{sanderson_unlocking_2025,sanderson_early_2024}, dynamo generation is possible. The two key factors controlling the \Rem are the lengthscale of convection \citep[assumed to be the radius of the liquid portion of the core;][]{sanderson_unlocking_2025} and the buoyancy flux at the CMB, which includes contributions from both core cooling and core solidification \citep[Equation 36;][]{sanderson_unlocking_2025}. Increasing either the convective lengthscale or the buoyancy flux increases the \Remp.

We made minor changes to the model relative to \citet{sanderson_unlocking_2025} to allow us to account for planetesimal water content. Model runs were started from the point of differentiation rather than accretion, and the mantle solidus, liquidus, and viscosity were changed to include the effects of water in NAMs.

\subsection*{Model start point}
Unlike in \citet{sanderson_unlocking_2025}, in this study the model begins at the point of differentiation rather than accretion to avoid modelling complexities due to water-rock differentiation and silicate (de)hydration prior to metal-silicate differentiation \citep{fu_fate_2014,castillo-rogez_origin_2017,trinh_slow_2023}. These processes will affect the time between accretion and differentiation, but since the differentiation criteria (reaching the temperature of critical melt fraction) determines the conditions for the subsequent thermal evolution and dynamo generation, of primary interest here, omitting this time interval does not impact our results. Additionally, the time between accretion and differentiation is very short (0.5--4\,Ma) in comparison to the time between differentiation and complete core solidification ($>100$\,Ma) \citep{kruijer_great_2020,spitzer_nucleosynthetic_2021}. This approach is similar to \citet{neumann_recurrent_2024}, who neglect the thermal effect of water prior to differentiation for planetesimals that form cores. 

The mean $\rm^{182}Hf/^{182}\rm W$ differentiation age of the NC grouped irons ($2.2\pm0.5$\,Ma after CAI formation) is older than that of the CC grouped irons \citep[$3.0\pm0.4$\,Ma after CAI formation;][]{hellmann_hf-w_2024}. However, iron meteorites from both reservoirs span a range of differentiation ages from 0.5--4.5\,Ma after CAI formation \citep{spitzer_nucleosynthetic_2021,hellmann_hf-w_2024,spitzer_comparison_2025}. Therefore, we adopted a differentiation time of 2\,Ma after CAI formation for both endmembers. Differentiation time has minimal effect on dynamo generation and does not alter the contrasting dynamo histories predicted for different endmembers (\ref{fig:tdiff}). This is because post-differentiation thermal evolution depends on the peak temperature of the mantle and mantle cooling rate. Differentiation time alters the abundance of radiogenic $^{26}\rm Al$ (the heat source for planetesimal thermal evolution) remaining after differentiation and its possible heating effect. However, due to the short half-life of $^{26}\rm Al$, radiogenic heating cannot delay mantle cooling beyond $\sim$5\,Ma after CAI formation \citep{sanderson_unlocking_2025}.

\subsection*{Effect of mantle water content on the mantle solidus and liquidus}
To capture the effect of water on mantle melting, we replaced the constant, dry solidus and liquidus values in \citet{sanderson_unlocking_2025} with the parametrisation presented by \citet{katz_new_2003} for the hydrated mantle solidus, $T_{\mm{m,s}}$ and liquidus, $T_{\mm{m,l}}$ converted to K,
\begin{equation}
    T_{\mm{m,s}} = A_1+A_2P+A_3P^2-\Delta T(X_{\mm{w}}), 
\label{eq:Tms}\end{equation} and
 \begin{equation}
    T_{\mm{m,l}}=C_1+C_2P+C_3P^2-\Delta T(X_{\mm{w}}).
\label{eq:Tml}\end{equation}
$P$ is the pressure in GPa, $A_1=1358.7\rm\,K$, $A_2=132.9\,\rm\,K GPa^{-1}$, $A_3=-5.1\,\rm\,K GPa^{-2}$, $C_1=2053\rm\,K$, $C_2=45\,\rm\,K GPa^{-1}$, $C_3=-2.0\,\rm\,KGPa^{-2}$. $\Delta T(X_{\mm{w}})$ is the temperature reduction for a water fraction, \Xwp, in wt\% in the melt,
\begin{equation}
    \Delta T(X_{\mm{w}})=K\left(\frac{X_{\mm{w}}}{D_{\mm{w,K}}+F(1-D_{\mm{w,K}})}\right)^\gamma,
\label{eq:dtwater}\end{equation}
where $D_{\mm{w,K}}=0.01$ is the bulk distribution coefficient of water between the solid and melt for this solidus parametrisation, $F$ is the melt fraction, and $K=43\rm\,K\,wt\,\%^{-\gamma}$, and $\gamma=0.75$ are constants \citep{katz_new_2003}. We assume the melt remains homogeneously mixed with the solid so the degree of melting and melt fraction are equivalent ($F=0$ at the solidus and $F=1$ at the liquidus). 

The dry solidus description from \citet{katz_new_2003} has a larger temperature difference between the solidus and liquidus compared to \citet{sanderson_early_2024,sanderson_unlocking_2025}. Apart from enabling lower initial core sulfur contents to be run in the model, the solidus parametrisation has minimal effect on the timing of thermal evolution and dynamo generation (Figure \ref{fig:solidus-comparison}). For further discussion of the solidus, see Section \ref{supp-solidus}.

\subsection*{Effect of mantle water content on the mantle viscosity}\label{met-eta}
The mantle viscosity is inversely proportional to the water content of the solid phase, $C_H^{\mm{sol}}$ \citep{hirth_water_1996} and the constant prefactor in the viscosity law is an order of magnitude higher than for dry silicate \citep[$A_0$;][]{keller_volatiles_2017}. Therefore, in planetesimals with water in NAMs, the first two pieces in the viscosity law \citep[Equations 1a and 1b in][]{sanderson_unlocking_2025} are multiplied by a factor of $\frac{10}{C_{H}^{\mm{sol}}}$. For a partially molten system, the water content of the solid phase depends on the melt fraction, $\phi$, and the partition coefficient of water between the solid and the melt, $D_{\mm{w}} = \frac{C_H^{\mm{sol}}}{C_H^{\mm{melt}}}$. \begin{equation}
    C_H^{\mm{sol}} = C_H^{\mm{tot}}\left(1-\phi + \frac{\phi}{D_{\mm{w}}}\right)^{-1}.
\end{equation} $C_H^{\mm{tot}}= 2\times10^4X_{\mm{w}}\frac{M_{\mm{min}}}{M_{H_2O}}$ is the total water content of the system in units of number of H atoms per million Si atoms, \Xw is the water concentration in wt \%, and $M_{\mm{min}}$ and $M_{H_2O}$ are the molar mass of the chosen mineral and water, respectively. In this model, we use the molar mass for KLB-1 peridotite ($M_{\mm{min}}=55.1$), which is a close natural analogue to upper mantle compositions \citep{davis_composition_2009} and matches the mantle composition in \citet{katz_new_2003}. We adopt the partition coefficient for peridotite at 1\,GPa, $D_{\mm{w}}=0.006$ from \citet{hirschmann_dehydration_2009}. Although this partition coefficient is for pressures 2-3 times higher than in planetesimal mantles, the pressure dependence of the partition coefficient is weak \citep{hirschmann_dehydration_2009}. Note, the value of this partition coefficient is different to the partition coefficient used to calculate the effect of water on the mantle solidus and liquidus ($D_{\mm{w,K}}$, Equation \ref{eq:dtwater}). For the mantle solidus, we use the same value $D_{\mm{w,K}}=0.01$ as in \citet{katz_new_2003}, because the constants in Equations \ref{eq:Tms}--\ref{eq:dtwater} have been calibrated against experimental data using this value. Water reduces viscosity, so for any given temperature, the viscosity of a planetesimal with water in its mantle will be less than or equal to that of a dry planetesimal. For our chosen constants, the $\frac{10}{C_H^{\mm{sol}}}$ modification to the viscosity law reduces viscosity for water contents $\geq0.0136$\,wt\% (Section \ref{supp-xw}). Therefore, water contents below this value were considered effectively dry.

The other parameters in the viscosity law change minimally for the water contents in NAMs adopted here \citep{mei_influence_2002,lesher_chapter_2015} and have smaller effects on dynamo generation than reference viscosity \citep{sanderson_early_2024}. Therefore, we used the constant values from \citet{sanderson_early_2024} for these parameters for all values of mantle water content. 

\subsection*{Parameter variation and planetesimal endmembers}
Firstly, we performed a parameter space exploration across a range of core radius fractions, \rcrp, and mantle water content in NAMs, \Xw (see below and \ref{tab:params}). Subsequently, we explored three endmember scenarios in detail based on possible planetesimal formation conditions and possible amounts of water retained after differentiation. Endmember 1 represented dry planetesimals that were reduced by nebula gas ($\frac{r_c}{r}=0.7,\: X_w=0$\,wt\%). Endmember 2 represented planetesimals that accreted a small fraction of water-ice and degassed all their water during differentiation ($\frac{r_c}{r}=0.5,\: X_w=0$\,wt\%). Endmember 3 represented planetesimals that accreted a lot of water-ice and retained some water in the mantle after differentiation ($\frac{r_c}{r}=0.4,\: X_w=0.07$\,wt\%). We ran models for 100-500\,km radius planetesimals with lower and upper limits set by the minimum size required for dynamo generation and the largest-surviving asteroids at the present day, respectively. All other parameter values were the same as Table 1 in \citet{sanderson_unlocking_2025} and the constant values in \citet{sanderson_early_2024}. In particular, we held differentiation time and initial core sulfur content constant. The justifications for our parameter ranges and sensitivity tests for differentiation time and initial core sulfur content are presented below.

\subsubsection*{Fractional core radius}\label{met-rcr}
In the parameter space exploration of \rcrp, we ran models for \rcrp=0.1--0.9 to investigate the widest possible range of fractional core radii. For the reducing gas endmember (E1), we used the core radius fraction predicted for an EH chondrite starting composition at IW-5 to the nearest one decimal place (0.7) based on the results of \citet{suer_formation_2025} (see Section \ref{supp-rcr}).  For the water-ice poor (E2) and water-ice rich (E3) endmembers, we used the average fractional core radii to the nearest one decimal place for NC (0.5) and CC (0.4) iron meteorites, respectively. We calculated these fractional core radii from the core mass fractions presented in \citet{spitzer_comparison_2025} (see Section \ref{supp-rcr}).

\subsubsection*{Water content in NAMs} 
Hydrated silicate minerals are unstable at the high temperatures ($\sim$1600\,K) reached by differentiated planetesimals \citep{lichtenberg_water_2019,trinh_slow_2023} and liquid water will have rapidly escaped as the planetesimal heated up \citep{fu_fate_2014}. Therefore, we assume the only water retained in the planetesimal is in small quantities in NAMs. The quantity of this water, \Xwp, can be estimated from measurements of water in NAMs in meteorites \citep{peterson_h_2023,newcombe_degassing_2023,rider-stokes_evidence_2024,harries_upper_2023,peterson_reconstruction_2025}. Petrological evidence suggests that the meteorites in which NAMs have been measured crystallised from a melt \citep{newcombe_degassing_2023}. Therefore, these measurements can be combined with the degree of melting on the parent body and water solid-melt partition coefficients to find the original water concentration in NAMs on the parent body \citep{newcombe_degassing_2023,peterson_reconstruction_2025}. We explored the range 0--700\,pm\,wt (\Xw =0--0.07\,wt\%) with the upper limit from \citet{newcombe_degassing_2023}, because their values account for water partitioning into multiple minerals and include measurements from both NC and CC achondrites. It is important to note that the 700\,ppm\,wt value from \citet{newcombe_degassing_2023} is also an experimental upper limit and the real values could be up to two orders of magnitude lower, because their measurements were at the detection limit. Since we want to explore the most extreme possibilities of water loss and retention, it is appropriate for us to use this upper limit. The concentration of water in NAMs is assumed to be constant throughout a model run. The concentrations of water in NAMs are too low for volatile exsolution to drive eruption; more oxidised melts (CM- and CV-like) are more dense than their overlying crusts \citep{fu_fate_2014}; and the low crustal permeability will inhibit melt transport via porous flow. Altogether, these factors will prevent melt from reaching the surface and water from within the melt degassing after differentiation. 

\subsection*{Core light element content}\label{met-xs}
The light elements that are likely to be present in planetesimal cores are S, O, C, Si and P \citep{pommier_joint_2020,bromiley_geochemical_2023,bromiley_effects_2026,suer_formation_2025}. For \fox $>$ IW-4, S will be the dominant light element \citep{suer_formation_2025}. Since Fe-FeS alloys have better constrained thermal and physical properties than iron alloys with multiple light elements \citep{pommier_joint_2020}, we model the core as having an Fe-FeS composition. This aligns with previous studies on compositional dynamo generation \citep[e.g.,][]{nimmo_energetics_2009,scheinberg_core_2016}. Sulfur is predicted to become more siderophile with increasing oxygen fugacity \citep{suer_formation_2025,bercovici_effects_2022} but there is no clear trend between core sulfur content in iron meteorite parent bodies and oxygen fugacity (Figure \ref{fig:fo2-rcr}). Additionally, planetesimal initial composition \citep{bercovici_effects_2022,suer_formation_2025}, immiscibility \citep{bromiley_effects_2026}, multi-stage differentiation \citep{grewal_protracted_2025}, and sulfur degassing during differentiation \citep{hirschmann_early_2021} can all alter core sulfur contents compared to the values expected from their bulk compositions. Therefore, we used the same initial core sulfur content (23\,wt\%) across the parameter space exploration and for all endmembers. Separately, we ran models with a range of initial core sulfur contents (23--33\,wt\%) for 300\,km radius planetesimals for each endmember to test the effect of this assumption on our results (\ref{fig:B-xs}).

The minimum initial core sulfur content we can input in our model is the lowest sulfur content that can be molten at differentiation. For a planetesimal with a solidus (and hence critical melt fraction) lowered by the presence of 0.07\,wt\% water, the lowest possible initial core sulfur content is 23\,wt\%, which is larger than the estimated core composition from almost all iron meteorites. We adopted this value across all end members because it was the value closest to the iron meteorite data that was within the allowed range of input values for all water contents.

\backmatter

\bmhead{Supplementary information}
A supplementary pdf with additional text and figures is available online. 

\bmhead{Acknowledgements}
HRS acknowledges funding on a Natural Environment Research Council studentship NE/S007474/1, an Exonian Graduate Scholarship from Exeter College, University of Oxford and funding from the Research Council of Norway through the Centres of Excellence funding scheme, project number 332523 (PHAB). JFJB acknowledges funding from the UKRI Research Frontier Guarantee program EP/Y014375/1. This study uses the Scientific colour maps imola and lajolla \citep{crameri_scientific_2023} to present the data in a visually uniform way that is accessible for all readers \citep{crameri_misuse_2020}.

\bmhead{Data Availability}
The dynamo and thermal evolution model and the parameter files required to recreate the results in this paper are publicly available at this https://zenodo.org/records/20527507 Github repository.

\bmhead{Declaration of Competing interests}
The authors declare that they have no known competing financial interests or personal relationships that could have appeared to influence the work reported in this paper.

\bmhead{Author contributions}
\textbf{Hannah R. Sanderson:} Conceptualization, Methodology, Software, Writing - Original Draft, Writing - Reviewing \& Editing. \textbf{James F. J. Bryson:} Conceptualization, Writing - Review \& Editing, Supervision. \textbf{Claire I. O. Nichols:} Conceptualization, Writing - Review \& Editing, Supervision.

\bibliography{nccc}

\begin{appendices}
\setcounter{table}{0}
\setcounter{equation}{0}
\setcounter{figure}{0}
\renewcommand{\thetable}{Extended Data Table \arabic{table}}
\renewcommand{\thefigure}{Extended Data Figure \arabic{figure}}
\section*{Extended Data}\label{edata}


\begin{table}
    \centering
    \begin{tabular}{|c|c|c|}\hline
        Parameter & Symbol & Value(s)  \\\hline
        \multicolumn{3}{|c|}{Constant parameters}\\\hline
        Initial core sulfur content & \Xs & 23\,wt\%   \\
        Differentiation time & $t_{\mm{diff}}$ & 2\,Ma after CAI formation \\
         Critical melt fraction & $\phi_C$ & 0.5 \\\hline
        \multicolumn{3}{|c|}{Variable parameters}\\\hline
        Fractional core radius & \rcr & 0.1--0.9  \\ 
        Water content in NAMs & \Xw & 0,0.02--0.07\,wt\% \\
        Differentiation time & $t_{\mm{diff}}$ & 0.5--4.5\,Ma after CAI formation \\
        Initial core sulfur content & \Xs & 23--33\,wt\% \\\hline
    \end{tabular}
    \caption{Constant and variable parameter values for parameter space exploration. Any parameters not stated are the same as those in Table 1 in \citet{sanderson_unlocking_2025} and the constant values in Table 1 in \citet{sanderson_early_2024}. Differentiation time and initial core sulfur content have both constant and variable parameter values for the main parameter space exploration (Figures \ref{fig:regime300}, \ref{fig:mag-main}, \ref{fig:regime100}, and \ref{fig:regime500}) and sensitivity tests (Figures \ref{fig:tdiff} and \ref{fig:B-xs}), respectively.}
    \label{tab:params}
\end{table}

\begin{figure}
    \centering
    \includegraphics[width=1\linewidth]{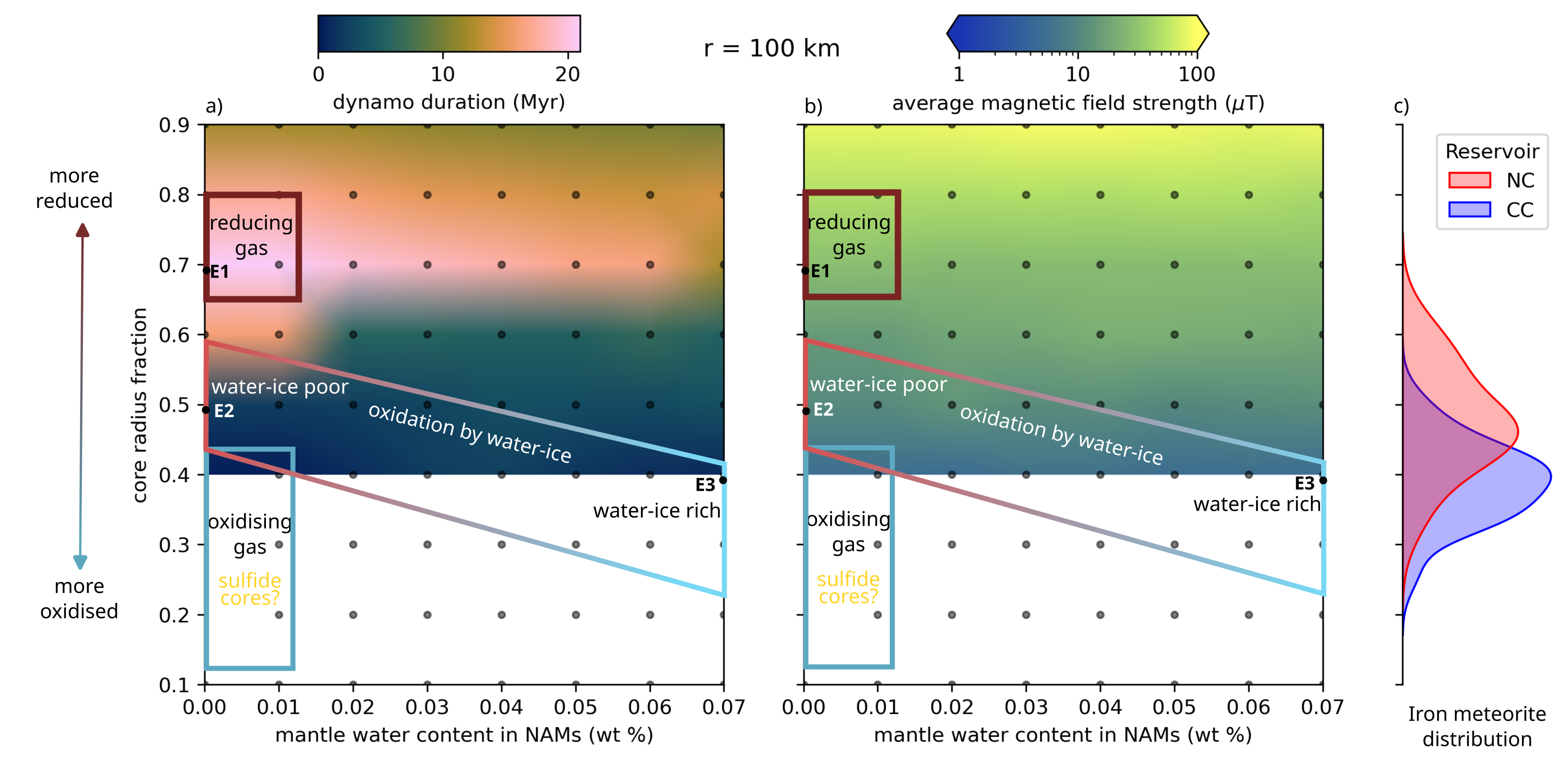}
    \caption{Same as Figure \ref{fig:regime300} but for 100\,km radius planetesimals. The range of values on the colourbar is the same as Figure \ref{fig:regime300} for magnetic field strength but different for total dynamo duration.}
    \label{fig:regime100}
\end{figure}

\begin{figure}
    \centering
    \includegraphics[width=1\linewidth]{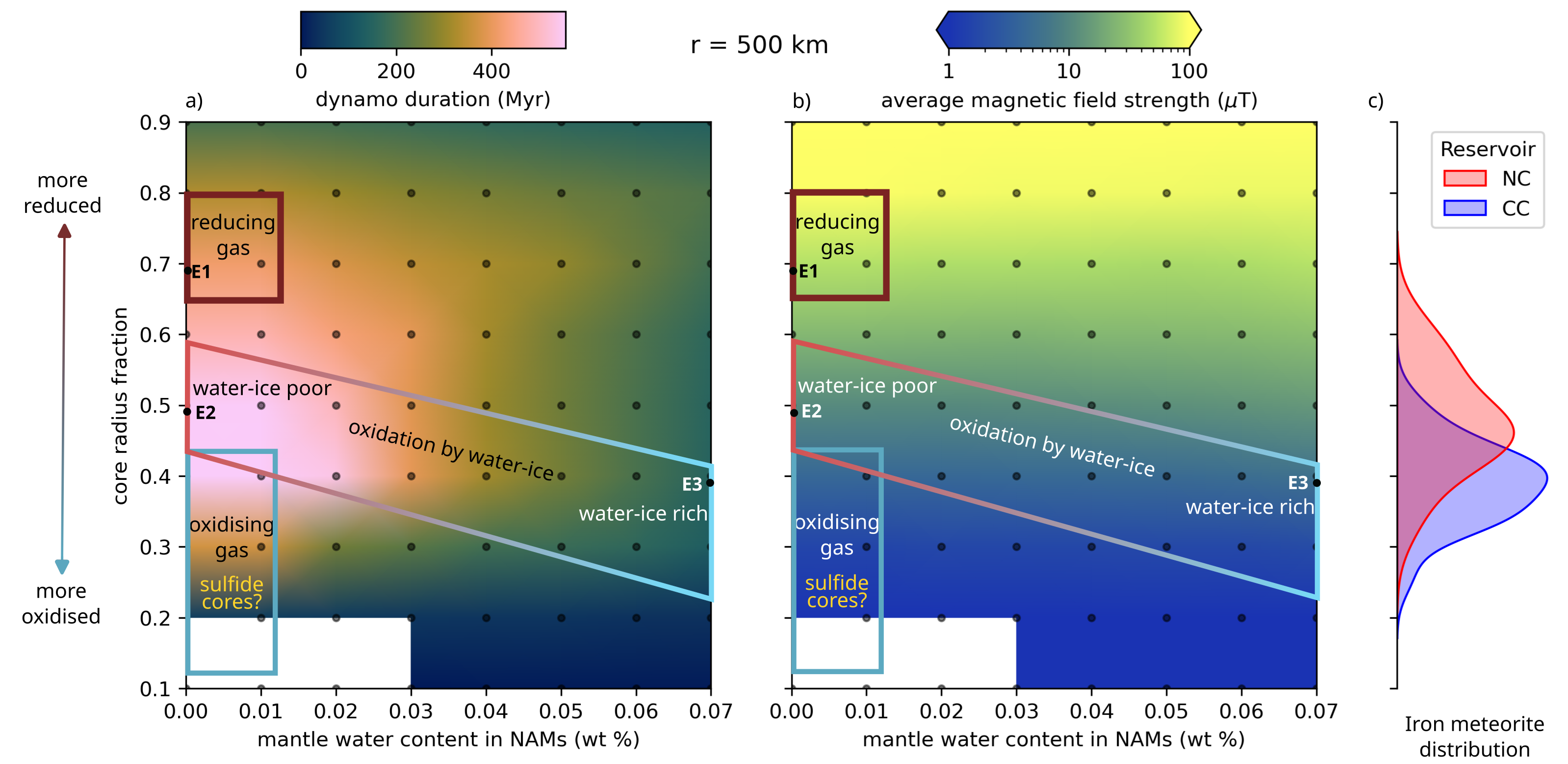}
    \caption{Same as Figure \ref{fig:regime300} but for 500\,km radius planetesimals. The range of values on the colourbar is the same as Figure \ref{fig:regime300} for magnetic field strength but different for total dynamo duration.}
    \label{fig:regime500}
\end{figure}

\begin{figure}
    \centering
    \includegraphics[width=1\linewidth]{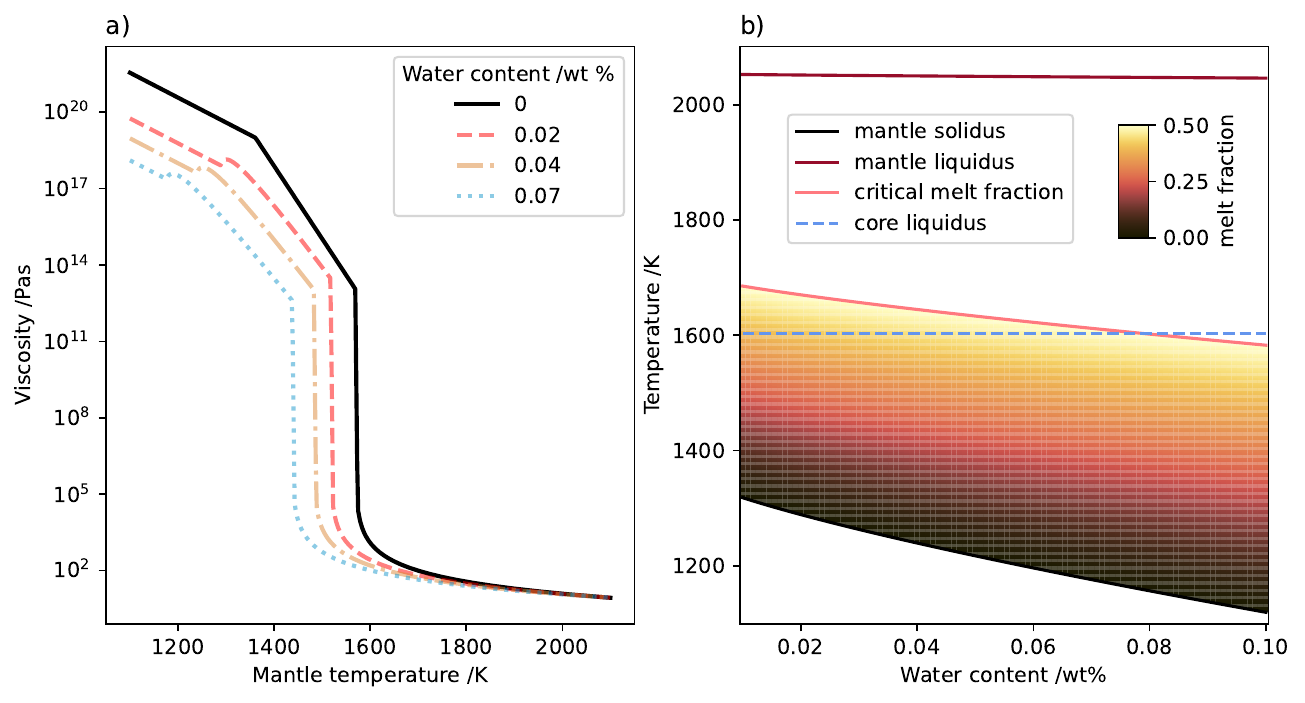}
    \caption{a) Viscosity profiles for a range of water contents assuming a dry reference viscosity of $10^{19}$\,Pas, $D_{\mm{w}}=6\times10^{-3}$ and a KLB1 peridotite composition. Below the solidus, wet materials have lower viscosities. At the solidus, wet materials have a slight increase in viscosity, because the water begins to move into the melt and the viscosity of the solid phase increases. The viscosity then decreases due to the increasing melt fraction with increasing temperature. b) Mantle solidus, $T_{\mm{m,s}}$, liquidus, $T_{\mm{m,l}}$, and temperature of the critical melt fraction, $T_{\phi_C}$, as a function of water content based on the solidus and liquidus parametrisation from \citet{katz_new_2003}. The critical melt fraction is assumed to be 0.5. The core liquidus, $T_{\mm{c,l}}$, is for a 300\,km radius planetesimal with 23\,wt\% sulfur in its core. }
    \label{fig:solidus-eta}
\end{figure}
\begin{figure}
    \centering
\includegraphics[width=0.8\linewidth]{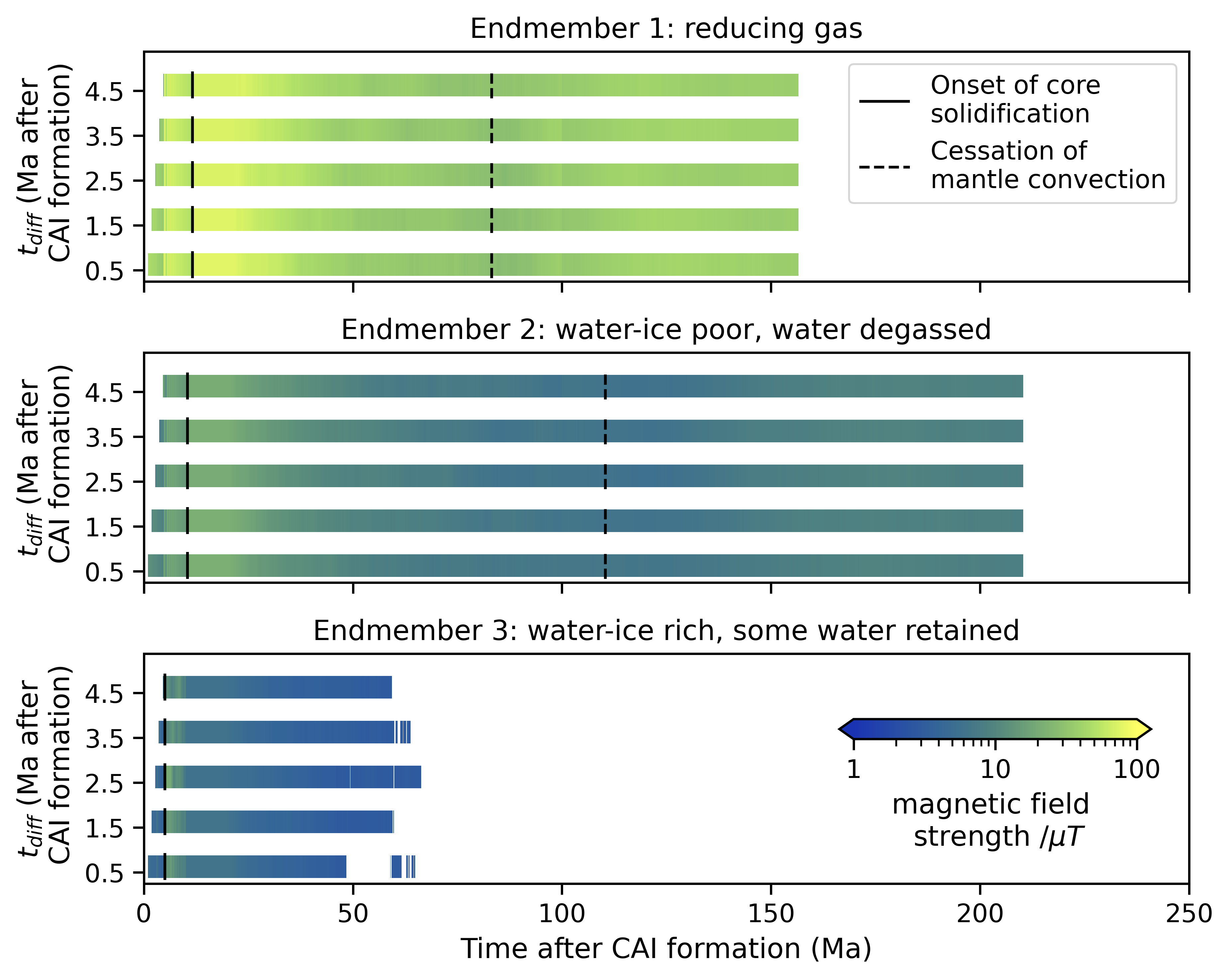}
    \caption{Dynamo strength and duration for the three planetesimal endmembers in Figure \ref{fig:mag-main} for a range of differentiation times. Filled bars indicate periods when a dynamo is active and the colour of the bar indicates dipole magnetic field strength at the surface. The black, vertical lines indicate the onset of core solidification (solid) and the cessation of mantle convection (dashed). Differentiation time has no effect on the long term dynamo generation. Endmember 1: dry planetesimals reduced by nebula gas ($\frac{r_c}{r}=0.7,\: X_w=$0\,wt\%). Endmember 2: planetesimals that accreted a small fraction of water-ice that degassed all their water during differentiation ($\frac{r_c}{r}=0.5,\: X_w=$0\,wt\%). Endmember 3: planetesimals that accreted a lot of water-ice and retained some water in the mantle after differentiation ($\frac{r_c}{r}=0.4,\: X_w=$0.07\,wt\%).}
    \label{fig:tdiff}
\end{figure}

\clearpage

\end{appendices}
\setcounter{table}{0}
\setcounter{section}{0}
\setcounter{subsection}{0}
\setcounter{subsubsection}{0}
\setcounter{equation}{0}
\setcounter{figure}{0}
\renewcommand{\thetable}{S\arabic{table}}
\renewcommand{\thesection}{S\arabic{section}}
\renewcommand{\theequation}{S\arabic{equation}}
\renewcommand{\thefigure}{S\arabic{figure}}

\section*{Dynamo generation reveals redox conditions during formation of differentiated planetesimals --- Supplementary Information}
\addtocontents{toc}{\protect\setcounter{tocdepth}{3}} 
\tableofcontents
\section{Additional background}
\subsection{Previous attempts to determine planetesimal oxidation state and water content}
Planetesimal accretion times can be calculated from $\rm^{182}Hf-^{182}W$ differentiation ages \citep{spitzer_nucleosynthetic_2021,hellmann_hf-w_2024,spitzer_comparison_2025} by using thermal evolution models to calculate the time delay between accretion and differentiation \citep{spitzer_nucleosynthetic_2021,bryson_collective_2026}. While some NC planetesimals differentiated earlier than CC planetesimals \citep{spitzer_nucleosynthetic_2021,spitzer_comparison_2025}, these models predict that they had similar accretion times \citep{spitzer_nucleosynthetic_2021,bryson_collective_2026}. This delay in differentiation is instead thought to be due to differences in the ratio of $\rm^{26}Al$-free water-ice to $\rm^{26}Al$-bearing silicates and refractory phases within the accreted material changing the abundance of $\rm^{26}Al$, and hence internal heating rate. These similar accretion times support either the `different condensation lines’ or `NC sub-reservoirs’ hypotheses. For accretion times, there are uncertainties due to the choice of Hf/W ratio of the precursor used in calculating differentiation ages \citep{hellmann_hf-w_2024} and the choice of model assumptions, particularly relating to hydrothermal convection and the differentiation mechanism \citep{spitzer_nucleosynthetic_2021,bryson_collective_2026}.

The redox states of differentiated planetesimals can be calculated from elemental abundances in magmatic iron meteorites, which are remnants of planetesimal cores \citep{hilton_chemical_2022,grewal_accretion_2024,spitzer_comparison_2025}. Redox state is quantified by oxygen fugacity, \foxp. Estimates of oxygen fugacity are also sensitive to many assumptions, including the behaviour of sulfur during fractional crystallisation, the highly siderophile element content of the chondritic precursor, and correlations between different element compositions  \citep{grewal_accretion_2024,spitzer_comparison_2025}.

Measurements of water contents in NAMs are challenging due to detection limits of instruments, uncertainties in partition coefficients, and terrestrial contamination \citep{newcombe_degassing_2023,harries_upper_2023,peterson_h_2023,stephant_hydrogen_2023,rider-stokes_evidence_2024}. As a result, many studies of water in NAMs in achondrites predict a wide range of parent body water contents for both NC \citep[e.g., 0.0007--0.02\,wt\%;][]{peterson_reconstruction_2025} and CC \citep[e.g., 0.0003--0.07\,wt\%;][]{newcombe_degassing_2023} bodies and studies on the same meteorite group may have contradictory results \citep[e.g., angrites;][]{sarafian_early_2017,deligny_origin_2021,rider-stokes_evidence_2024}. Our result that NC planetesimals degassed efficiently strengthens the argument that high achondrite water contents in NAMs likely result from a combination of poorly understood partitioning behaviour, terrestrial contamination, and values below instrument detection limits \citep{newcombe_degassing_2023,peterson_h_2023,peterson_h-poor_2024,rider-stokes_evidence_2024,peterson_reconstruction_2025}.

\section{Additional model details}

\subsection{Core sulfur contents}
We do not include changes in core sulfur content with oxygen fugacity due to the lack of trend in the iron meteorite data (see \nameref{met-xs}). In the main text, we adopt the same core sulfur content for all endmembers. Here, we present a series of model runs across a range of initial core sulfur contents (23--33\,wt\%) to justify this assumption.

\begin{figure}
    \centering
\includegraphics[width=1\linewidth]{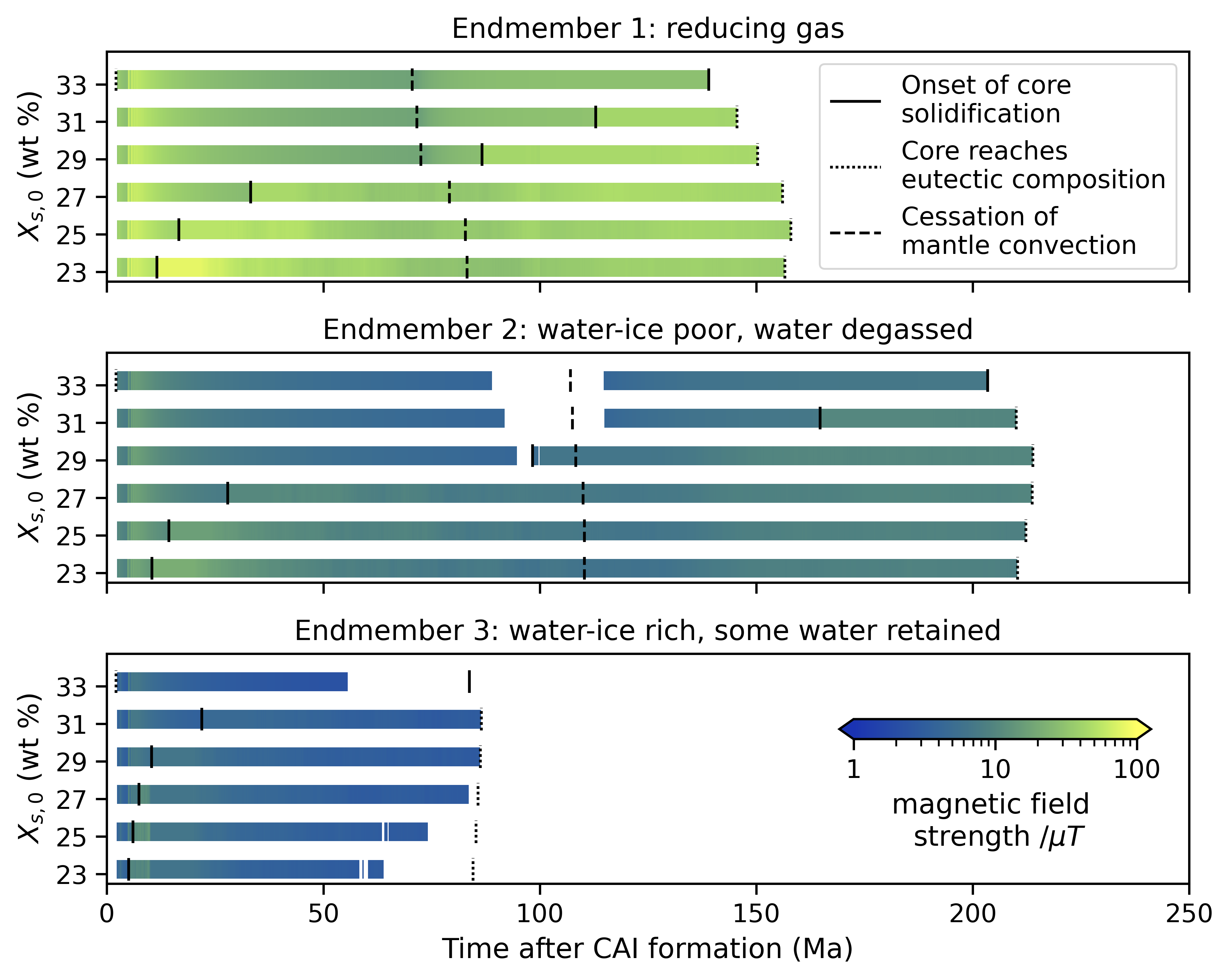}
    \caption{Dynamo strength and duration for the three planetesimal endmembers in Figure \ref{fig:mag-main} as a function of initial core sulfur content, \Xsp. The minimum initial core sulfur content on the vertical axis is set by the peak temperature at differentiation for Endmember 3. Filled bars indicate periods when a dynamo is active and the colour of the bar indicates dipole magnetic field strength at the surface. The black vertical lines indicate the onset of core solidification (solid), when the core reaches the eutectic composition (dotted), and the cessation of mantle convection (dashed). Endmember 1: dry planetesimals reduced by nebula gas ($\frac{r_c}{r}=0.7,\: X_w=$0\,wt\%). Endmember 2: planetesimals that accreted a small fraction of water-ice that degassed all their water during differentiation ($\frac{r_c}{r}=0.5,\: X_w=$0\,wt\%). Endmember 3: planetesimals that accreted a lot of water-ice and retained some water in the mantle after differentiation ($\frac{r_c}{r}=0.4,\: X_w=$0.07\,wt\%).}
    \label{fig:B-xs}
\end{figure}

In our models, the core liquidus temperature depends on the core sulfur content. As a result, variations in initial core sulfur content change the onset time of core solidification, altering when compositional convection can begin to contribute to dynamo generation. However, varying initial core sulfur content does not remove the contrast between planetesimal endmembers (Figure \ref{fig:B-xs}). Increased initial core sulfur contents in Endmember 2 can lead to a gap in dynamo generation just before and after the cessation of mantle convection because core solidification has not yet begun and there is no compositional buoyancy to drive the dynamo. This gap could lead to degeneracy in null measurements in this time period between Endmember 2 and 3. Multiple magnetic measurements from a single meteorite group where non-zero paleointensities follow null paleointensities, like those for the Main Group pallasites \citep{nichols_time-resolved_2021}, could resolve this degeneracy. Additionally, iron meteorite compositions suggest these high initial core sulfur contents are unlikely to occur in meteorite parent bodies (Section \ref{met-xs}). Instead the initial core sulfur content is likely to be at the bottom of or below the range explored here, although how lower sulfur content cores differentiate is still an area of active research \citep[e.g.][]{grewal_protracted_2025}. Since varying the initial core sulfur content does not alter the main differences in dynamo generation between planetesimal endmembers, the unknown the initial core sulfur content of the different endmembers does not preclude predictions of the dynamo histories of these bodies.

\subsubsection{Sulfide cores}
For bodies formed above the iron-w{\"u}stite buffer with sufficient sulfur, such as those formed between the tar and water-ice line or in the outermost CC reservoir, their cores are predicted to be super-eutectic in the Fe-FeS system and crystallise FeS rather than Fe \citep{crossley_parent_2023,bercovici_effects_2022,crossley_percolative_2025}. We currently have no large, sulfide meteorites from these possible cores, probably due to the mechanical weakness of sulfides, which prevent them from reaching to Earth's surface intact \citep{crossley_parent_2023,mccoy_deciphering_2022}.  The material properties of sulfide cores are too uncertain \citep{ruckriemen_top-down_2018} to predict dynamo generation histories for these bodies. Although we show their location in Figure \ref{fig:regime300} for comparison to other planetesimal endmembers, the dynamo duration and average magnetic strength displayed is not valid for these compositions and we do not investigate them in this study. Further research on the thermal and electrical properties of sulfides are required to predict magnetic histories for bodies with sulfide cores \citep{ruckriemen_top-down_2018}. 

\subsection{Minimum water content for a wet planetesimal}\label{supp-xw}
For my modification to the prefactor in the viscosity law (Section \nameref{met-eta}) to reduce the viscosity of the wet planetesimals requires 
\begin{equation}
    1 \geq \frac{10}{C_H^{\mm{sol}}}.
\end{equation}
After substituting for $C_H^{\mm{sol}}$ in terms of $C_H^{\mm{tot}}$ and $C_H^{\mm{tot}}$ in terms of \Xwp, this becomes
\begin{equation}
   1 \geq \frac{10(1-\phi+\frac{\phi}{D_{\mm{w}}})}{2\times10^4X_{\mm{w}}\frac{M_{\mm{min}}}{M_{H_2O}}}
\end{equation}
For $\phi=0.5$, $D_{\mm{w}}=0.006$, and KLB-1 peridotite mineralogy ($M_{\mm{min}}=55.1$), $X_{\mm{w}}$ must exceed 0.0136\,wt\%. Water contents below this value are so low they do not affect the viscosity and the dry viscosity law should be used.

\subsection{Differences in solidus parametrisation}\label{supp-solidus}

\begin{figure}
    \centering
    \includegraphics[width=0.9\linewidth]{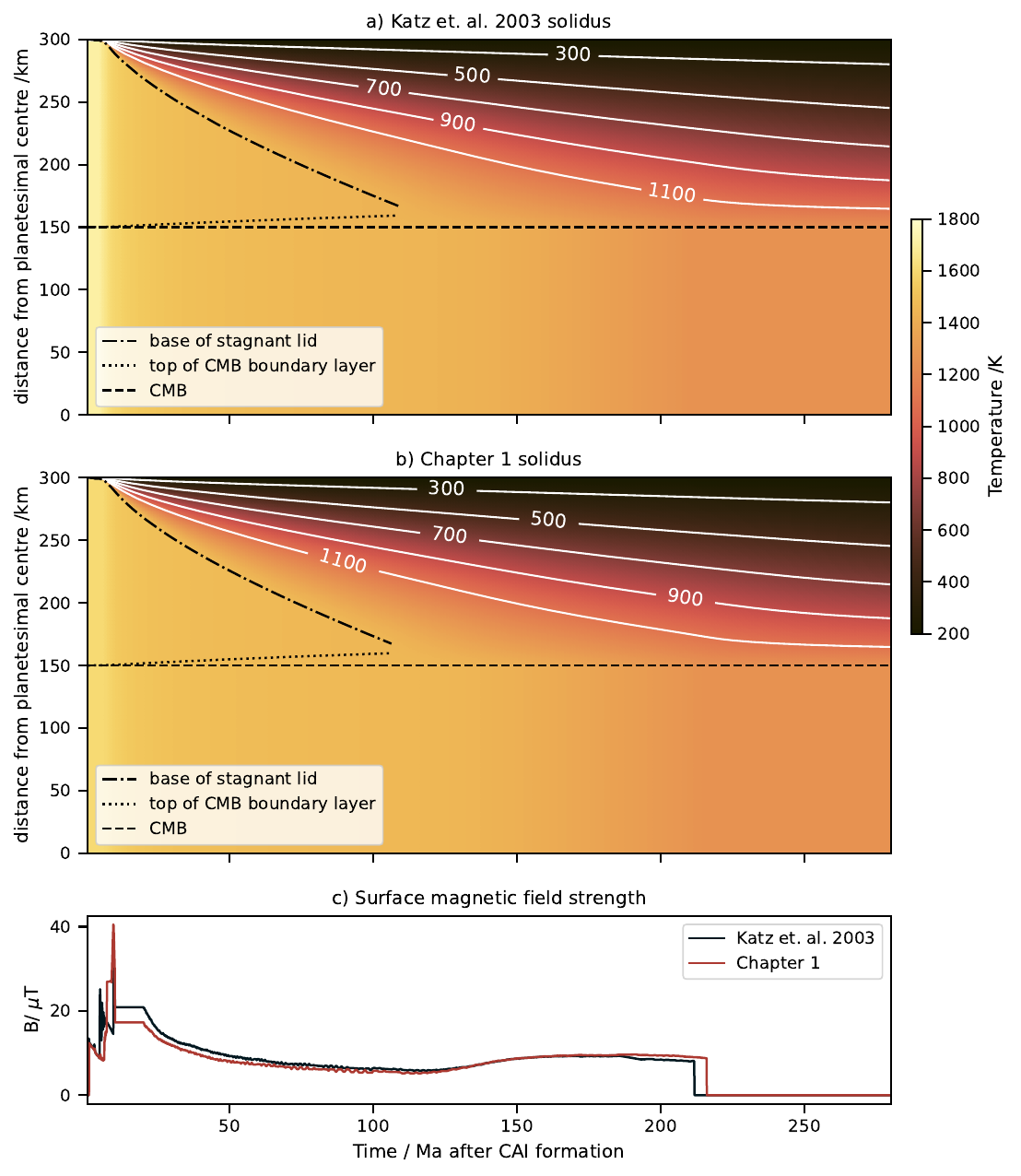}
    \caption[Thermal evolution of two identical, 300\,km radius, dry planetesimals using the \citet{katz_new_2003} solidus and the \citet{sanderson_unlocking_2025} solidus]{Thermal evolution of two identical, 300\,km radius, dry planetesimals using a) the \citet{katz_new_2003} solidus and b) the \citet{sanderson_unlocking_2025} solidus. The peak temperature is higher for the \citet{katz_new_2003} solidus, but otherwise the thermal evolutions are very similar. c) magnetic field strength, $B$, as a function of time for the \citet{katz_new_2003} solidus parametrisation (black trace) and the constant values used in \citet{sanderson_unlocking_2025} (brown trace). The spike in the brown trace at early times comes from the earlier onset of core solidification due to the lower differentiation temperature.}
    \label{fig:solidus-comparison}
\end{figure}

The larger temperature difference between the solidus and liquidus in the dry solidus description from \citet{katz_new_2003} compared to \citet{sanderson_early_2024,sanderson_unlocking_2025} raises the temperature of the critical melt fraction. This enables lower \Xs values to be run in the model (4.44\,wt\% compared to 22.63\,wt\% for dry, 300\,km radius planetesimals with critical melt fractions equal to 0.3). Other than the initial core sulfur content, the solidus parametrisation has minimal effect on the timing of thermal evolution and dynamo generation (Figure \ref{fig:solidus-comparison}). Identical, 300\,km radius, dry planetesimals with differing solidii have $<5$\% difference in their times for the cessation of mantle convection, dynamo generation, and core solidification. 

The iron concentration of silicate minerals strongly affects their solidii \citep{kiefer_effects_2015}. This results in a $\sim40$\,K difference in dry solidus temperature between the parametrisation of \citet{katz_new_2003} for mantle peridotite \citep[Mg\#90;][]{davis_composition_2009}, and that of \citet{sanderson_unlocking_2025} based on CC chondrite Allende \citep[Mg\#69;][]{agee_pressure-temperature_1995}. Iron contents in planetesimal mantles will be lower than those in chondrites due to movement of iron to the core during differentiation. Therefore, the composition assumed in \citet{katz_new_2003} is more appropriate for the iron content in differentiated planetesimals. We neglect differences in dry solidus temperature between the wet and dry planetesimal endmembers due to changing core fraction because addition of 0.07\,wt\% water lowers the solidus more (185\,K) than the addition of iron ($\leq40$\,K depending on composition). 

Silicate mineralogy also affects the solidus, because different minerals have different partition coefficients for water \citep{rider-stokes_evidence_2024}. We adopt the parametrisation of \citet{katz_new_2003}, because it enables easy quantification of the effect of mantle water content on the solidus. While the trend in solidus depression with increasing water content will be similar in planetesimal mantles, the exact temperature differences may vary due to mineralogical differences between mantle peridotite and planetesimal mantles, and mantle heterogeneity.

\subsection{Effect of water content and core size on mantle heat transfer}
At the beginning of a planetesimal's thermal evolution, mantle viscosity is low due to the presence of melt and the mantle transfers heat via stagnant lid convection. In stagnant lid convection, conductive boundary layers are present at the surface and CMB \citep[for a schematic see][]{sanderson_unlocking_2025}. As the mantle cools, the amount of melt decreases, mantle viscosity increases, and these convective boundary layers thicken. When the combined boundary layer thickness is equal to the thickness of the mantle, mantle convection ends and mantle heat transport becomes purely conductive. When the mantle has a low viscosity, the boundary layers are thin and there is a high CMB heat flux, which supports dynamo generation. As the layers thicken, the CMB heat flux decreases, which can lead to the cessation of dynamo generation around the time of the cessation of convection. Once mantle convection has ceased, the conductive gradient at the CMB gradually steepens which can lead to the onset of a second period of dynamo generation \citep{sanderson_unlocking_2025}. 

\begin{figure}
    \centering
    \includegraphics[width=1\linewidth]{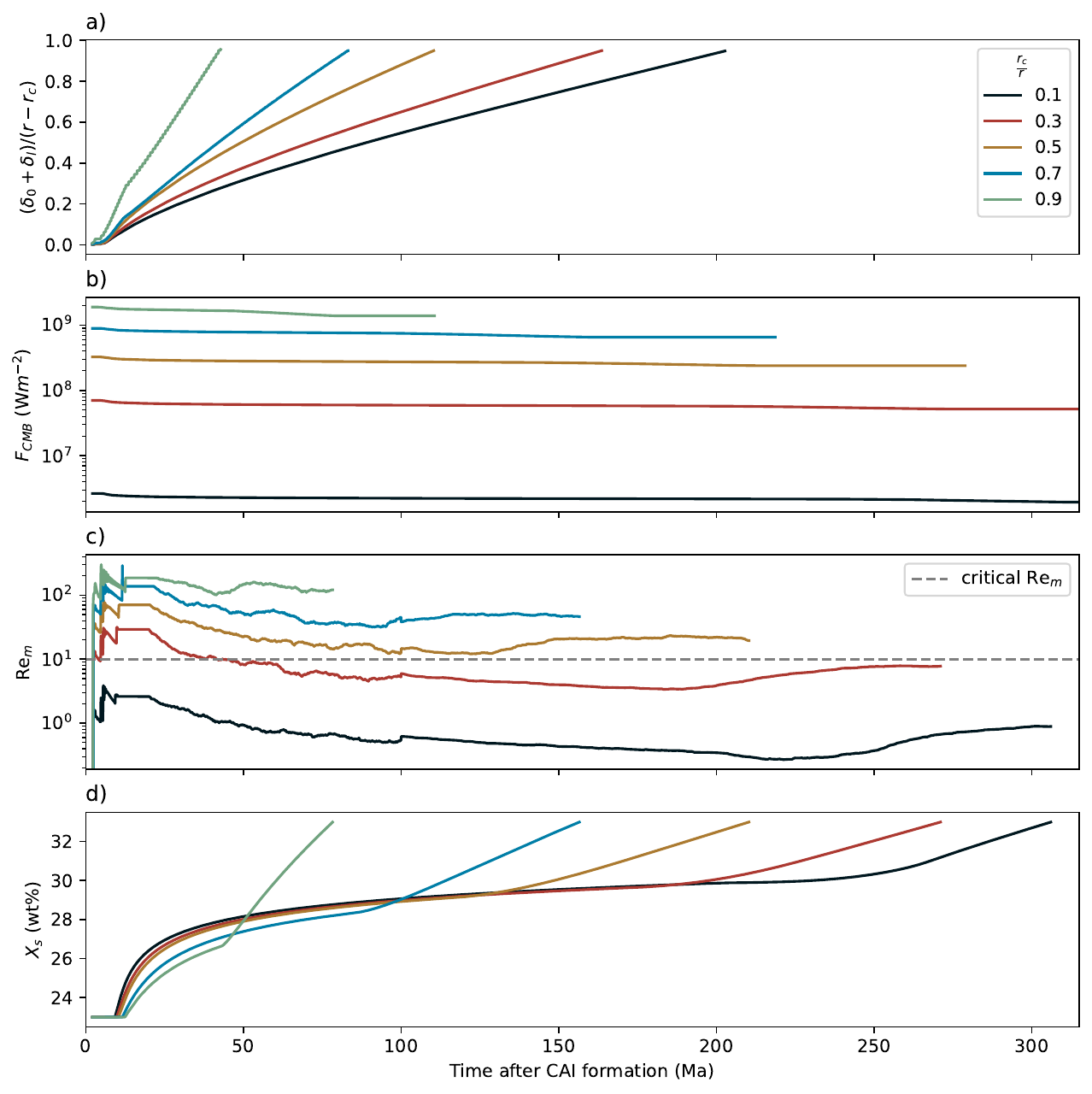}
    \caption[Convective boundary layer thickness, CMB heat flux, magnetic Reynolds number, and core sulfur content as a function of time for a 300\,km radius planetesimal with a range of core radius fractions]{a) Total convective boundary layer thickness as a fraction of mantle thickness, $(\delta_0+\delta_l)/(r-r_\mm{c})$, b) CMB heat flux, $F_{\rm CMB}$, c) magnetic Reynolds number, $Re_m$, and d) core sulfur content, $X_{\mm{S}}$, as a function of time for a 300\,km radius planetesimal with a range of fractional core radii, \rcrp, and no water in NAMs. Time evolution of each parameter in each panel is truncated a) when mantle convection ceases, b) when the core completely solidifies, c) and d) when the core reaches the eutectic composition. The time the final core reaches the eutectic composition sets the upper limit on the horizontal axis range. }
    \label{fig:fluxes-rcr}
\end{figure}

Increasing mantle water content lowers mantle viscosity. This increases the time for which the CMB boundary layer is thin and rapid core cooling is possible. This brings forward the onset and completion of core solidification. Increasing core radius fraction, decreases mantle thickness. Convection ceases earlier in thinner mantles, because less cooling is required before the convective boundary layer thickness is equal to the mantle thickness (Figure \ref{fig:fluxes-rcr}). Once mantle convection ceases, the shorter distance for conduction between the CMB and the surface compared to planetesimals with smaller core radius fractions also increases the rate of core cooling. 

\section{Additional information for figures}
\subsection{Fractional core radii as a function of oxygen fugacity}\label{supp-rcr}
\subsubsection{Iron meteorites}
The fractional core radius for each iron meteorite (Table \ref{tab:rcr}) was calculated from the core mass fractions and core sulfur contents from \citet{spitzer_comparison_2025}. The mantle was assumed to have a constant density of 3000$\,\rm kgm^{-3}$ and the core was assumed to have a constant density that depended on core sulfur content according to the relationship used in \citet{sanderson_unlocking_2025}. The mean fractional core radii for the NC and CC iron meteorites are $0.48\pm0.07$ and $0.39\pm0.06$, respectively. The non-magmatic ungrouped iron meteorites in \citet{spitzer_comparison_2025} were excluded from the mean. To align with the values used in the parameter space exploration, we used values of 0.5 and 0.4 for the water-ice poor, efficient degassing (E2) and the water-ice rich, water retained (E3) endmembers, respectively.

\subsubsection{Chondritic starting compositions}
The core radius fraction for H and EH chondritic starting compositions as a function of oxygen fugacity was calculated from the core mass fractions and core densities presented in \citet{suer_formation_2025} assuming a constant mantle density of 3000$\,\rm kgm^{-3}$. 

\subsection{Boundaries between disk reservoirs in Figure \ref{fig:regime300}}\label{supp-redox}
The boundaries for the 3 reservoirs as shown in Figures \ref{fig:regime300}, \ref{fig:regime100}, \ref{fig:regime500} are justified as follows:
\begin{itemize}
    \item Reducing gas --- the core radius fraction range is for an EH core density and core mass fraction at $<$IW-5 (\rcrp=0.67) from \citet{suer_formation_2025} and Mercury \citep[\rcrp=0.8;][]{hauck_ii_curious_2013}, which is thought to have formed in the same region as enstatie chondrites. This reservoir was very close to the Sun so there was negligible water in these planetesimals (upper limit of 0.0136\,wt\% in this model, Section \ref{supp-xw}).
    \item  Oxidising gas --- the range of core radius fractions is based on the volume \% of sulfides in R chondrites \citep[0.2--9.2\% sulfides, \rcr=0.12--0.45;][]{bischoff_rumuruti_2011}, assuming that all sulfide goes into the core during differentiation. This reservoir is hypothesised to have been between the water-ice line and the tar line so while these planetesimals would have been surrounded by an oxidising gas there would be no water in the mantles of these bodies.
    \item Oxidation by water ice --- within these reservoirs oxidation state is hypothesised to vary due to variations in amount of water-ice accreted so this spans the entire range of mantle water contents to account for possible water retention. The four corners were based on the maximum and minimum core radius fractions from the iron meteorite data \citep{grewal_accretion_2024,spitzer_comparison_2025} for IW$<-2$ and IW$>-1$ for the 0\,wt\% and 0.07\,wt\% water endmembers respectively. (0, 0.6) Zacatecas (1792); (0, 0.45) IIAB; (0.07, 0.26) Tishomingo; (0.07, 0.43) Grand Rapids.
\end{itemize}

\section{Additional figures and tables}
\clearpage
\begin{table}[]
    \centering
    \begin{tabular}{|c|c|c|c|c|c|}\hline
Meteorite & $f\rm O_2$ ($\Delta IW$) & $\delta f\rm O_2$ ($\Delta IW$) & \makecell{Core sulfur\\content (wt\%)}& \makecell{Core mass\\fraction (wt\%)} & \makecell{Fractional\\core radius} \\\hline
\multicolumn{6}{|c|}{\textbf{NC}}\\\hline
IC & -1.90 & 0.00 & 18 & 25 & 0.54 \\
IIAB & -2.50 & 0.50 & 17 & 27 & 0.56 \\
IIIAB & -1.65 & 0.05 & 12 & 19 & 0.48 \\
IIIE & -1.55 & 0.05 & 12 & 18 & 0.47 \\
IVA & -1.85 & 0.15 & 6 & 19 & 0.46 \\
Reed City & -1.70 & 0.10 & 12 & 18 & 0.47 \\
Zacatecas (1792) & -2.55 & 0.15 & 24 & 31 & 0.61 \\
Santiago Papasquiero & -1.95 & 0.15 & 2 & 17 & 0.44 \\
Washington County & -1.50 & 0.10 & 8 & 14 & 0.42 \\
EET 83230 & -1.40 & 0.10 & 2 & 9 & 0.35 \\\hline
\multicolumn{6}{|c|}{\textbf{CC}}\\\hline
IIC & -1.40 & 0.00 & 8 & 13 & 0.41 \\
IID & -1.40 & 0.00 & 10 & 13 & 0.41 \\
IIF & -1.30 & 0.10 & 13 & 13 & 0.42 \\
IIIF & -2.20 & 0.50 & 5 & 20 & 0.47 \\
IVB & -1.30 & 0.10 & 1 & 8 & 0.33 \\
SBT & -1.15 & 0.05 & 7 & 7 & 0.33 \\
Babbs Mill (Troost's Iron) & -1.10 & 0.10 & 8 & 8 & 0.34 \\
Grand Rapids & -1.55 & 0.15 & 8 & 15 & 0.43 \\
Pi$\rm\tilde{n}$on & -1.30 & 0.10 & 3 & 8 & 0.34 \\
Tucson & -1.65 & 0.15 & 2 & 13 & 0.40 \\
Mbosi & -1.65 & 0.15 & 7 & 15 & 0.43 \\
New Baltimore & -2.05 & 0.15 & 12 & 22 & 0.50 \\
Nordheim & -1.50 & 0.10 & 2 & 11 & 0.37 \\
NWA 6932 & -1.30 & 0.10 & 10 & 11 & 0.39 \\
Tishomingo & -0.90 & 0.10 & 1 & 4 & 0.26 \\
ALHA 77255 & -1.45 & 0.15 & 1 & 10 & 0.36 \\
Guffey & -1.60 & 0.10 & 1 & 13 & 0.39 \\
Hammond & -1.60 & 0.10 & 12 & 17 & 0.46 \\
ILD 83500 & -1.10 & 0.10 & 8 & 8 & 0.34 \\
Illinois Gulch & -1.45 & 0.15 & 4 & 11 & 0.38 \\
La Caille & -1.55 & 0.15 & 8 & 15 & 0.43 \\\hline
    \end{tabular}
    \caption{Core sulfur contents, core mass fractions, oxygen fugacities from \citep{spitzer_comparison_2025} and fractional core radii (this study) for grouped and ungrouped iron meteorites. The oxygen fugacity presented here is the average of the oxygen fugacities calculated from Fe/Ni and Fe/Co and the error is half the difference between these values. The ungrouped non-magmatic iron meteorites from \citet{spitzer_comparison_2025} have been excluded.}
    \label{tab:rcr}
\end{table}

\begin{table}[]
    \centering
    \begin{tabular}{|c|c|c|c|c|} \hline
        Meteorite  & Group & Age (Ma after CAI formation) & Paleointensity ($\mu$T) & Reference \\\hline
Portales Valley & H chondrite & 105$\pm$25 & 16 & \citet{bryson_paleomagnetic_2019}\\
Bjurb{\"{o}}le & L/LL chondrite & 110$\pm$30 & 9 & \citet{shah_long-lived_2017} \\
Colomera & IIE iron & 97$\pm$10 & 36 & \citet{maurel_meteorite_2020} \\
Techado & IIE iron & 78$\pm$13 & 15 & \citet{maurel_meteorite_2020} \\
Miles & IIE iron & 159$\pm$9 & 33 & \citet{maurel_long-lived_2021}\\
Marjalahti & MG pallasite & 98$\pm$12 & 0 & \citet{nichols_time-resolved_2021} \\
Brenham & MG pallasite & 115$\pm$15 & 0 & \citet{nichols_time-resolved_2021}\\
Springwater & MG pallasite & 132$\pm$10 & 22$\pm$8 & \citet{nichols_time-resolved_2021}\\
Imilac & MG pallasite & 160$\pm$10 & 14 (IQR\,=\,16) & \citet{nichols_time-resolved_2021}\\
Esquel & MG pallasite & 190$\pm$28 & 10 (IQR\,=\,11)& \citet{nichols_time-resolved_2021}\\\hline
    \end{tabular}
    \caption{Paleomagnetic data used in Figure \ref{fig:mag-main}. For Imilac and Esquel, paleointensities are the median values when the uncertainty in paleofield direction is accounted for calculated by bootstrapping \citep{nichols_time-resolved_2021}. The interquartile range (IQR) obtained by the same method is given as a measure of spread for the paleointensity distribution due to uncertainty in paleofield direction. For Marjalahti and Brenham, their paleointensities are listed as 0, because their measured in paleointensities are consistent with a null field. The age ranges on the Main Group (MG) pallasites are the oldest and youngest times of remanence acquisition for inner core and outer core solidification models \citep{nichols_microstructural_2018}. The timing of remanence acquisition is either from radiometric dating (IIE irons, H chondrites) or thermal modelling (Main Group pallasites and L/LL chondrite). The ages from thermal models may change slightly for different modelling assumptions \citep{sanderson_early_2024}.}
    \label{tab:paleoint}
\end{table}

\begin{figure}
    \centering
\includegraphics[width=0.8\linewidth]{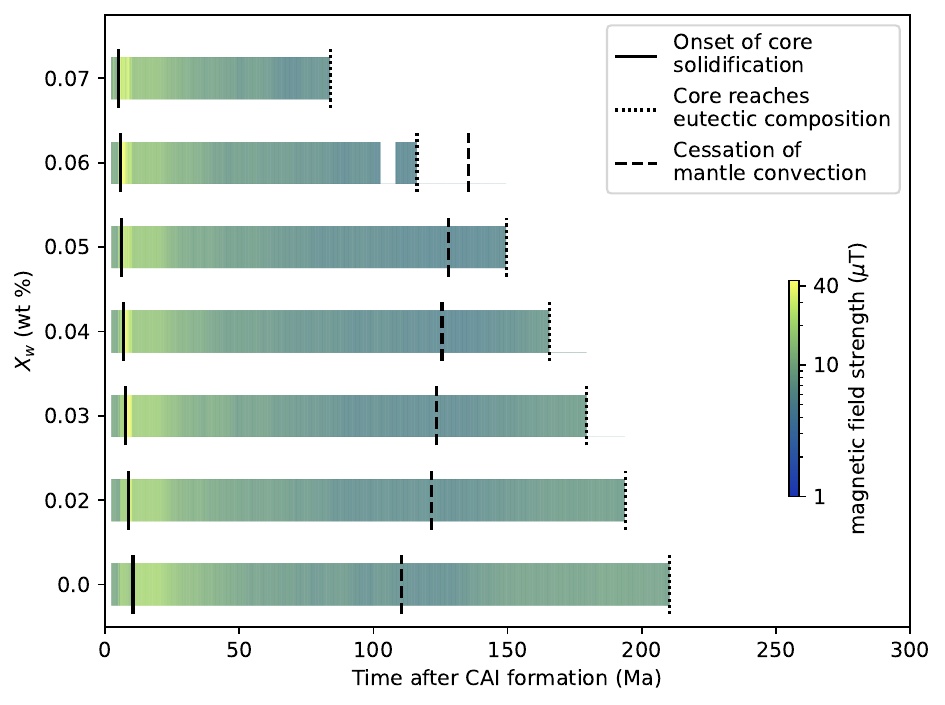}
    \caption{Dynamo strength and duration as a function of water content in nominally anhydrous minerals, \Xwp, for a 300\,km radius planetesimal with core radius fraction equal to 0.5. Filled bars indicate periods when a dynamo is active and the colour of the bar indicates dipole magnetic field strength at the surface. The black, vertical lines indicate the onset of core solidification (solid), when the core reaches the eutectic composition (dotted) and the cessation of mantle convection (dashed).}
    \label{fig:B-xw}
\end{figure}

\begin{figure}
    \centering
    \includegraphics[width=1\linewidth]{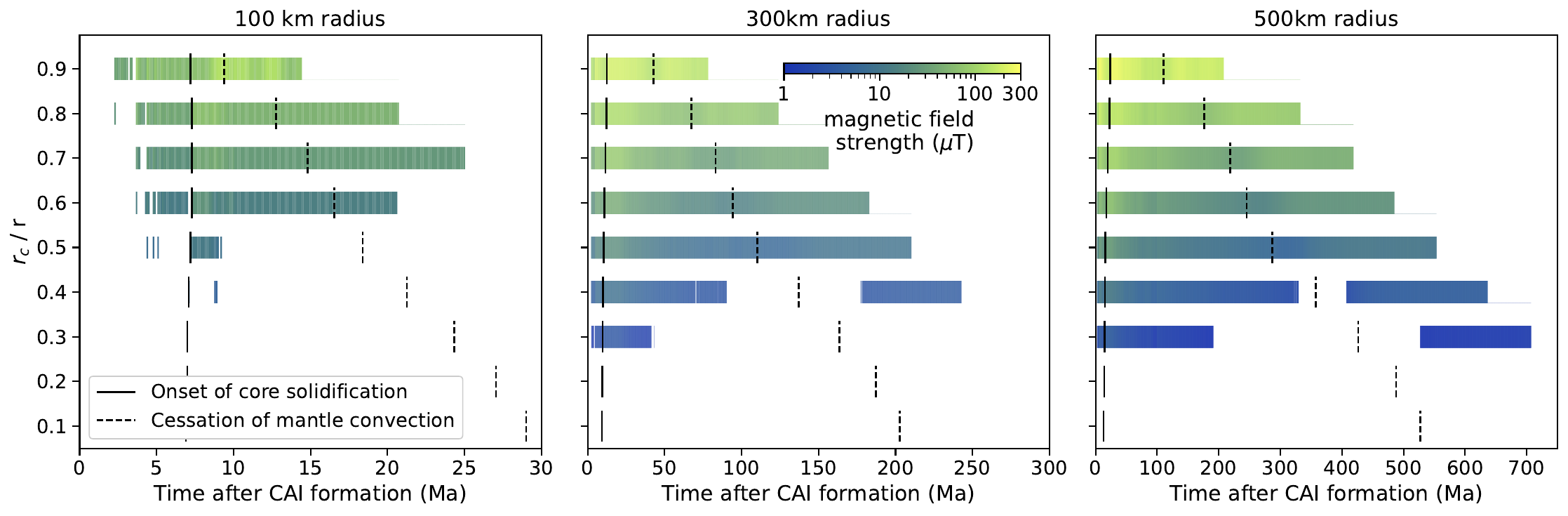}
    \caption{Dynamo strength and duration as a function of fractional core radius, \rcrp, for 100\,km, 300\,km and 500\,km radius planetesimals with no water in NAMs (\Xwp=0\,wt\%). Filled bars indicate periods when a dynamo is active and the colour of the bar indicates dipole magnetic field strength at the surface. The black, vertical lines indicate the onset of core solidification (solid) and the cessation of mantle convection (dashed).}
    \label{fig:B-rcr}
\end{figure}

\begin{figure}
    \centering
    \includegraphics[width=1\linewidth]{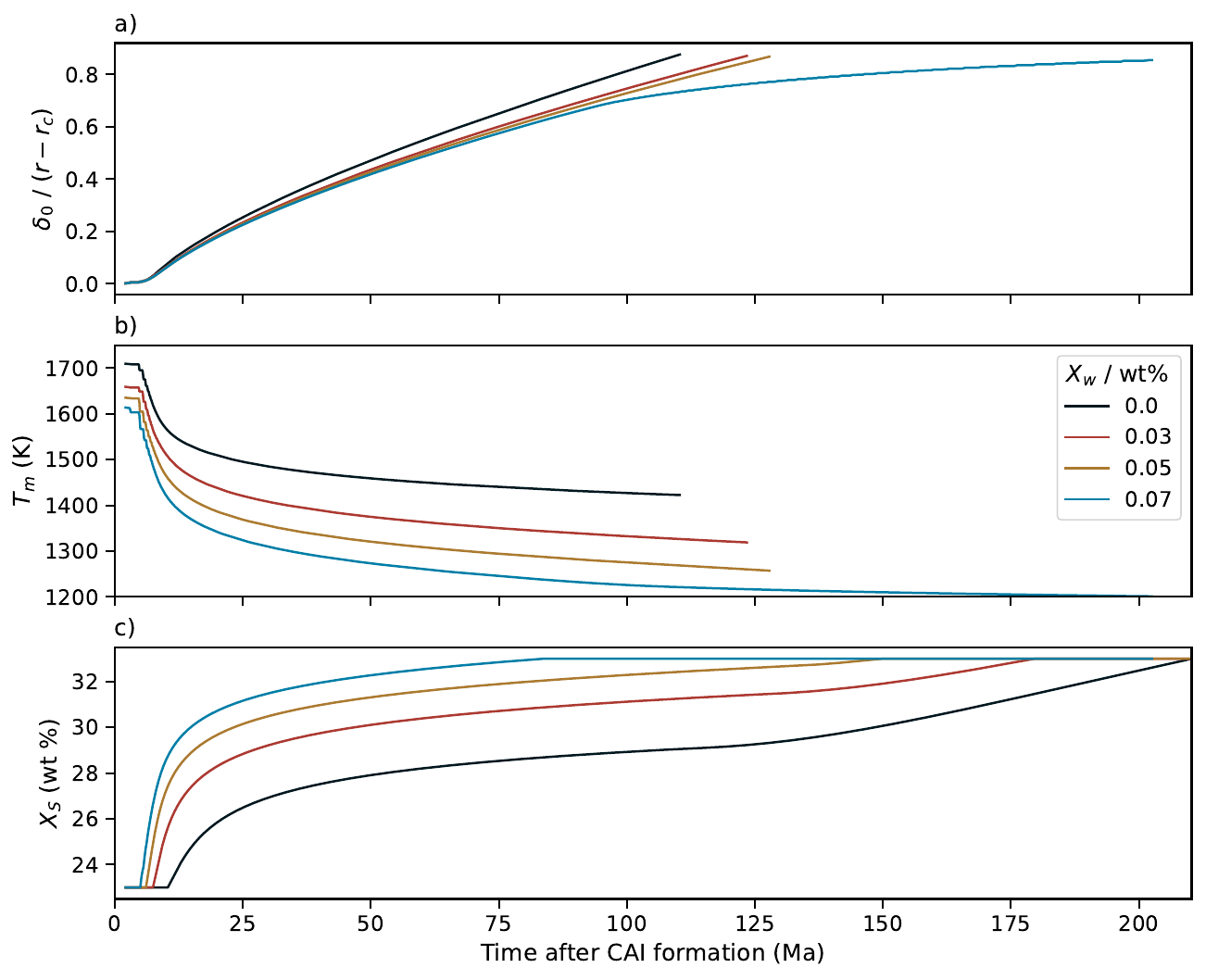}
    \caption[Stagnant lid thickness, convective mantle temperature, and core sulfur content as a function of time for a 300\,km radius planetesimal with a range of water contents in nominally anhydrous minerals]{a) Stagnant lid thickness as a fraction of mantle thickness, $\delta_0/(r-r_\mm{c})$, b) convective mantle temperature, $T_\mm{m}$, and c) core sulfur content, $X_{\mm{S}}$, as a function of time for a 300\,km radius planetesimal with a range of water contents in NAMs, \Xwp, and a core radius fraction of 0.5. Stagnant lid thickness and mantle convective temperature are plotted until the cessation of convection.}
    \label{fig:fluxes-xw}
\end{figure}

\begin{figure}
    \centering
    \includegraphics[width=1\linewidth]{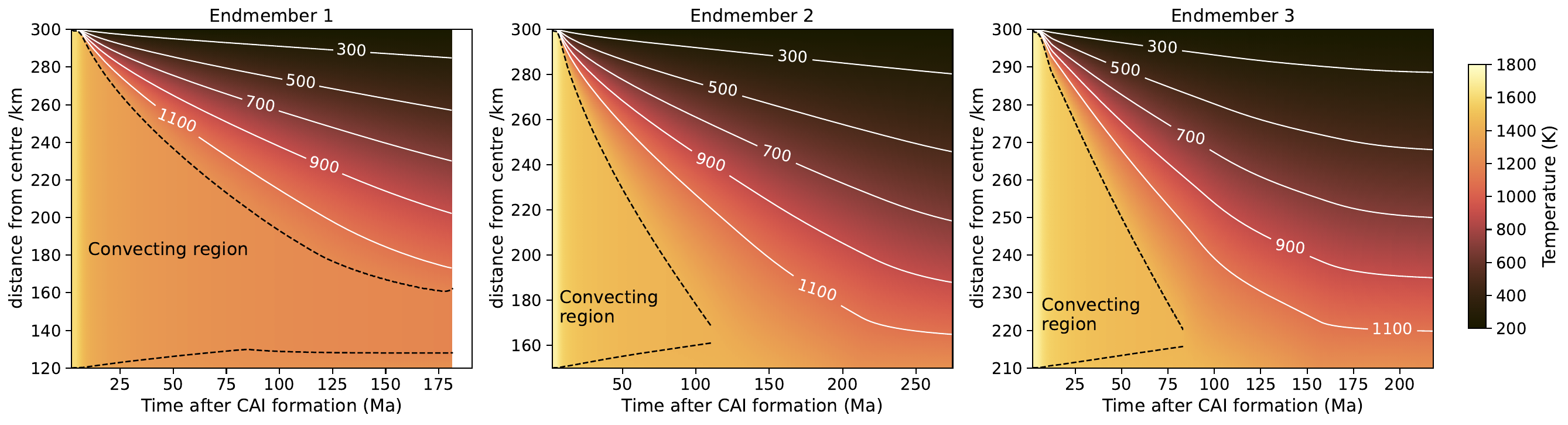}
    \caption[Interior temperature profiles for the three planetesimal endmembers]{Interior temperature profiles for the three planetesimal endmembers with radii of 300\,km. The region enclosed by the black dashed lines is convecting. The temperature contours span the range of thermochronometer closure temperatures in Figure \ref{fig:mag-main} and are all located in the conductive portion of the planetesimal. The horizontal axis limit is set by the solidification of the planetesimal core. Each panel has a different vertical axis range because of the difference in core size between the planetesimals. The horizontal axis ranges are set by the time taken for the core to solidify in each scenario. Endmember 1: dry planetesimals reduced by nebula gas ($\frac{r_c}{r}=0.7,\: X_w=$0\,wt\%). Endmember 2: planetesimals that accreted a small fraction of water-ice that degassed all their water during differentiation ($\frac{r_c}{r}=0.5,\: X_w=$0\,wt\%). Endmember 3: planetesimals that accreted a lot of water-ice and retained some water in the mantle after differentiation ($\frac{r_c}{r}=0.4,\: X_w=$0.07\,wt\%).}
    \label{fig:temp-profile}
\end{figure}

\end{document}